\documentclass[journal]{ieeetran}
\usepackage{cite}
\usepackage[pdftex]{graphicx}
\usepackage{epstopdf}
\usepackage{amsmath}
\usepackage{cases}
\usepackage[caption=false,font=footnotesize]{subfig}
\usepackage{fixltx2e}
\usepackage{amssymb}
\usepackage{mathtools}
\usepackage{bm}
\usepackage{multirow}
\usepackage{color}
\usepackage{algorithm}
\usepackage{algorithmic}
\usepackage{mathrsfs}
\usepackage{amsthm,amsmath,amssymb}
\usepackage{threeparttable}
\usepackage{url}
\usepackage{bbding}
\usepackage{wasysym}
\newcommand{\tabincell}[2]{\renewcommand\arraystretch{0.9}\begin{tabular}{@{}#1@{}}#2\end{tabular}}

\begin{document}

\title{Optimal Sizing and Pricing of Renewable Power to Ammonia Systems Considering the Limited Flexibility of Ammonia Synthesis}
\author{Zhipeng~Yu,
        Jin~Lin,~\IEEEmembership{Member,~IEEE},
        Feng~Liu,~\IEEEmembership{Senior Member,~IEEE},
        Jiarong~Li,
        Yuxuan~Zhao,
        Yonghua~Song,~\IEEEmembership{Fellow,~IEEE},
        Yanhua~Song, and
        XinZhen~Zhang
\thanks{Financial support came from the National Key R\&D Program of China (2021YFB4000504). \emph{(Corresponding author: Jin Lin)}
}
\thanks{Zhipeng Yu, Jin Lin, Feng Liu, Jiarong Li and Yuxuan Zhao are with the State Key Laboratory of Control and Simulation of Power Systems and Generation Equipment, Department of Electrical Engineering, Tsinghua University, Beijing 100087, China. And Jin Lin is also with Sichuan Energy Internet Reasearch Institute, Tsinghua University, Chengdu, 610213, China. (e-mail: linjin@tsinghua.edu.cn) %(e-mail:yuzp20@mails.tsinghua.edu.cn; linjin@tsinghua.edu.cn; lfeng@tsinghua.edu.cn; jr-li@mail.tsinghua.edu.cn)
}
\thanks{Yonghua Song is with the Department of Electrical and Computer Engineering, University of Macau, Macau 999078, China, and also with the Department of Electrical Engineering, Tsinghua University, Beijing 100087, China. %(e-mail: yhsong@tsinghua.edu.cn)
}
\thanks{Yanhua Song is with Shenzhen Edo Renewable Co., Ltd, Shenzhen, 518067, China.}
\thanks{Xinzhen Zhang is with Sichuan Energy Internet Reasearch Institute, Tsinghua University, Chengdu, 610213, China.}
}
\maketitle

\begin{abstract}
    Converting renewable energy into ammonia has been recognized as a promising way to realize ``green hydrogen substitution" in the chemical industry. However, renewable power to ammonia (RePtA) requires an essential investment in facilities to provide a buffer against the strong volatility of renewable energy and the limited flexibility of ammonia synthesis, which involves the three main stakeholders, namely, power, hydrogen, and ammonia. Therefore, the sizing and pricing of RePtA play a core role in balancing the interest demands of investors. This paper proposes an optimal sizing and pricing method for RePtA system planning. First, power to ammonia (P2A) is modeled as a flexible load, especially considering the limited flexibility of ammonia synthesis, which has been verified using real dynamic regulation data. Second, the multi-investor economic (MIE) model is established considering both external and internal trading modes. Then, a two-stage decomposed sizing and pricing method is proposed to solve the problem caused by the strong coupling of planning, operation, and trading, and information gap decision theory (IGDT) method is utilized to handle the uncertainty of renewable generation. Finally, real data from a real-life system in Inner Mongolia are utilized to verify the proposed approach. The results show that the system proposed has a yield of 8.15\%.
\end{abstract}


\vspace{-12pt}
\section{Introduction}
\label{sec:intro}
\subsection{Background and Motivation}
\IEEEPARstart{D}{ecarbonization} of the energy sector is becoming the goal of world development in the future \cite{papadis2020challenges}, which greatly increases the pressure of $\mathrm{CO_2}$ emission reduction in the chemical industry. Take China as an example, its annual output of ammonia ($\mathrm{NH_3}$) has reached 48 million tonnes and is heavily dependent on coal and natural gas, with an average annual $\mathrm{CO_2}$ emission of more than 145 Mt in 2015 \cite{zhang2019driving}. Moreover, among all traditional hydrogen-based chemical industries, ammonia has been the largest hydrogen downstream market in China \cite{palys2021renewable}. Therefore, there is great potential for \emph{green hydrogen substitution} in the ammonia industry.

Thus, a large-scale green hydrogen substitution mode in the ammonia industry is proposed as an on-grid renewable power to ammonia (RePtA) system, shown in Fig. \ref{fig:OperationTopu}. In this way, it can improve the local consumption ability of renewable energy on the one hand, and realize the $\mathrm{CO_2}$ emission reduction of the ammonia industry on the other hand.

\textcolor{black}{China's Inner Mongolia published the projects of wind-solar-hydrogen integrated systems in 2022, there are four projects selecting the technology of renewable power to ammonia \cite{Batou-Project-2022}. These RePtA systems need nearly about 1.3 GW installed capacity of wind turbines and PV, while producing over 300,000 tonnes ammonia per year. Rapidly increasing engineering demands require the method, to address system capacity optimization and business mode design, which are the motivations and main works of our research.}
%in the seven published projects
%\textcolor{blue}{Engineering background should be introduced clearly: wind-solar-hydrogen-ammonia integrated system! At the moment, connecting to hydrogen network or hydrogen supply chain is not the focus.}
%\footnotetext[1]{1 ${\rm{RMB}} \approx 0.1381\ \$ $(according to the exchange rate on 23th, Oct. 2022). }
%\footnotetext[2]{$1\ {\rm{Nm}}^3$ $\rm{H_2}\approx 11.2 \ \rm{kg}$ $\rm{H_2}$. }
%\let\thefootnote\relax\footnotetext{1 ${\rm{Nm}}^3$ $\rm{H_2}$ is equal to 11.2 $\rm{kg}$ $\rm{H_2}$. }
\vspace{-12pt}
\subsection{Literature Review}
However, there are still some problems in the planning of RePtA system. The literature review can be summarized from the following two aspects.

\textcolor{black}{\emph{1) Load modeling of P2A.} RePtA systems have two main types of loads, namely, electrolysis and ammonia synthesis.}

\textcolor{black}{Electrolyzers are the key devices in power to hydrogen (P2H) systems. For commercial applications, alkaline electrolysis has outstanding advantages in techno-economic \cite{schiebahn2015power}. For a single alkaline electrolyzer, the variable load range is 20\%$\sim$100\%, and the maximum ramping rate exceeds 20\% load/s \cite{li2021co,koponen2015review,ziems2012project,mehrtash2020enhanced,sanchez2018optimal,allman2018optimal}. Furthermore, for a multi-electrolyzer cluster system, with advanced control strategies, such as pressure control strategies \cite{qi2021pressure} and flexible startup and shutdown strategies \cite{qiu2022extended}, it is possible to expand the load range of the cluster system by 5\%$\sim$120\%. Therefore, P2H is regarded as a highly flexible load, especially in hourly resolution planning.}

\textcolor{black}{On the other hand, fewer studies discuss the flexibility of ammonia synthesis, due to the safety requirements for chemical production. \cite{beerbuhl2015combined} and \cite{hasan2019development} treat ammonia synthesis as a constant load without flexibility, while it is treated as a totally flexible load without any limitation in \cite{verleysen2020can,sanchez2018optimal}. Moreover, \cite{klyapovskiy2021optimal} and \cite{armijo2020flexible} summarize the parameters determining the flexibility of ammonia, including the variable load range and ramping limitation. The above is summarized in Table \ref{tab:difference_flexibility}.}

\textcolor{black}{Existing works, however, ignore the difference in flexibility between P2H and P2A. And the feasibility of dynamic regulation of ammonia synthesis and related practical limitations are lacking in further analysis.}

\emph{2) Business mode design. }Usually, generation companies mainly invest in power generation \cite{kong2019multi,sinha2008formulation}, part of which involves the investment in hydrogen production. Chemical or gas companies are more likely to invest in the storage and transmission of hydrogen \cite{kroniger2014hydrogen}, as well as in ammonia synthesis \cite{zhongming2020yara}, and some invest in hydrogen production. Therefore, taking hydrogen production as the boundary, its upstream and downstream regions belong to different investors.

At present, most of the existing works are conducted from the perspective of a single investor \cite{li2021co,palys2019novel,beerbuhl2015combined,varela2021modeling,klyapovskiy2021optimal,mehrtash2020enhanced,qi2021two,zheng2022incorporating}, with maximizing the revenue of the whole system as the goal. \cite{li2021co} proposes a co-planning model for wind to ammonia (WtA) and electric network (EN), by minimizing the total investment costs. \cite{klyapovskiy2021optimal} designs the optimal schedule with maximized benefits for a hydrogen-based energy management system, including electrolysis plants, hydrogen storage tanks, electric batteries, and hydrogen-consuming plants. \cite{zheng2022incorporating} proposes a multi-objective optimization framework to reveal optimal investment plans, maximizing both economic revenue and green hydrogen production.

These works, however, ignore the interest demands of different investors. %How to balance the interests of different investors in the RePtA system is still underexplored.
\textcolor{black}{How to design the business mode of the RePtA system, both external trading in the electricity and ammonia market (price-taker), and internal trading among different investors (price-maker), are still underexplored.}
\begin{table}[t]\scriptsize
  \renewcommand{\arraystretch}{2.5}
  \caption{Difference of Flexibility between P2H and P2A}
  \label{tab:difference_flexibility}
  \centering
  \vspace{-9pt}
  \scalebox{0.88}
  {
  \begin{tabular}{ccc}
  \hline \hline
  ${}$   &\tabincell{c}{Alkaline Electrolysis}
  &\tabincell{c}{Ammonia Synthesis}\\
  \hline
  Load variation range     &\tabincell{c}{20\%-100\% \cite{li2021co} \\5\%-150\% \cite{ziems2012project} \\30\%-100\% \cite{allman2018optimal}}
  &\tabincell{c}{50\%-100\% \cite{palys2019novel} \\60\%-105\% \cite{klyapovskiy2021optimal} \\75\%-110\% \cite{allman2018optimal}}  \\
  \hline
  Ramping up rates     &$+20\%$ load/s \cite{armijo2020flexible}      &\tabincell{c}{$+20\%$ load/h \cite{klyapovskiy2021optimal} \\$+10\%$ load/h \cite{palys2019novel}}  \\
  \hline
  Ramping down rates     &$-20\%$ load/s \cite{armijo2020flexible}      &\tabincell{c}{$-20\%$ load/h \cite{klyapovskiy2021optimal,palys2019novel}}  \\
  \hline
  Other limitations     &\tabincell{c}{None}      &\tabincell{c}{Change no more than once every 4 h \cite{palys2019novel} \\ Cold stop minimum duration is 48 h \cite{klyapovskiy2021optimal}}  \\
  \hline \hline
  \end{tabular}
  }
  \vspace{-12pt}
\end{table}

\vspace{-12pt}
\subsection{Contributions}
\textcolor{black}{To fill these gaps}, this paper first proposes a planned quasi-steady-state condition scheduling method for dynamic regulation of ammonia synthesis, trading off the flexibility and safety of ammonia synthesis, as shown in Fig. \ref{fig:Ammonia_Adjustment_Fig}. Then, a multi-investor economic (MIE) model is proposed, including the external and internal trading mode, as shown in Fig. \ref{fig:EconomicTopu}. Finally, a two-stage decomposed sizing and pricing method is proposed: \textcolor{black}{1) Maximum revenue is achieved in stage I, by confirming the optimal sizes of all facilities, considering the external trading in the electricity and ammonia market. And a robust model based on information gap decision theory (IGDT) is established, to handle the uncertainty of renewable generation. 2) Interest demands of different investors are balanced in stage II, by confirming the optimal inner electricity and hydrogen price, considering the internal trading among different investors.} %which avoids a large number of bilinear terms in form and the limited bilinear terms remaining are easy to handle.
%\textcolor{blue}{To fill these gaps}, this paper proposes a multi-investor economic model for the first time, that is, decoupling the economic analysis of different investors by the settlement of inner electricity and hydrogen prices, as shown in Fig. \ref{fig:EconomicTopu}. Then, a planned quasi-steady-state condition scheduling method for ammonia synthesis is proposed, which describes the variable load range, ramping limit and scheduling period at the same time, trading off the flexibility and safety of ammonia production. Finally, a two-stage decomposed sizing and pricing method is proposed, which avoids a large number of bilinear terms in form and in which the limited bilinear terms remaining are easy to handle. %\textcolor{blue}{Maximum revenue is achieved in stage I, considering outside trading in electricity and ammonia market, while interest demands of different investors are balanced in stage II, by confirming the optimal inner electricity and hydrogen price in the inside trading.}

Specifically, following contributions are made in this paper:

\textcolor{black}{1) Limited flexibility of P2A is modeled for the first time. Unlike P2H, the dynamic regulation process of ammonia synthesis in P2A can not be ignored in the hourly resolution planning, which is established as a 1-order transition process. The real data are used for parameter identification, and results show that the proposed model is accurate with tolerable error ($<3\%$). And, pressure and temperature of ammonia synthesis are changed in a safety range, satisfying the practical limitations.}
%Power to ammonia (P2A) is modeled as a limited flexible load for the first time.

\textcolor{black}{2) A multi-investor economic model is established to jointly consider the planning, operation, and trading of the RePtA system. Moreover, the IGDT method is utilized to handle the uncertainty of renewable generation. To circumvent the obstacle of large-scale bilinear terms arising from the model, we reveal that the sizing subproblem and the pricing subproblem can be decoupled in order. This finding enables a novel two-stage decomposed method to solve the model effectively.}
%\textcolor{blue}{2) Multi-investor economic model is established to describe the planning, operation and trading of RePtA system. And a two-stage decomposed sizing and pricing method is proposed to address the issues of large-scale bilinear terms. The proposed approach decouples the sizing and pricing in order, utilizes IGDT method to handle the uncertainty of renewable power, and avoids a large number of bilinear terms in form, the limited bilinear terms remaining are easy to handle.}

\textcolor{black}{3) The influence of ammonia flexibility on sizing and pricing is studied using real data. The results indicate that improving the flexibility of ammonia is essential to enhance the techno-economic benefits of the system, especially when the scheduling period of ammonia is less than one week.}

\begin{figure}[t]
  \centering
  \vspace{-12pt}
  \includegraphics[width=3.46in]{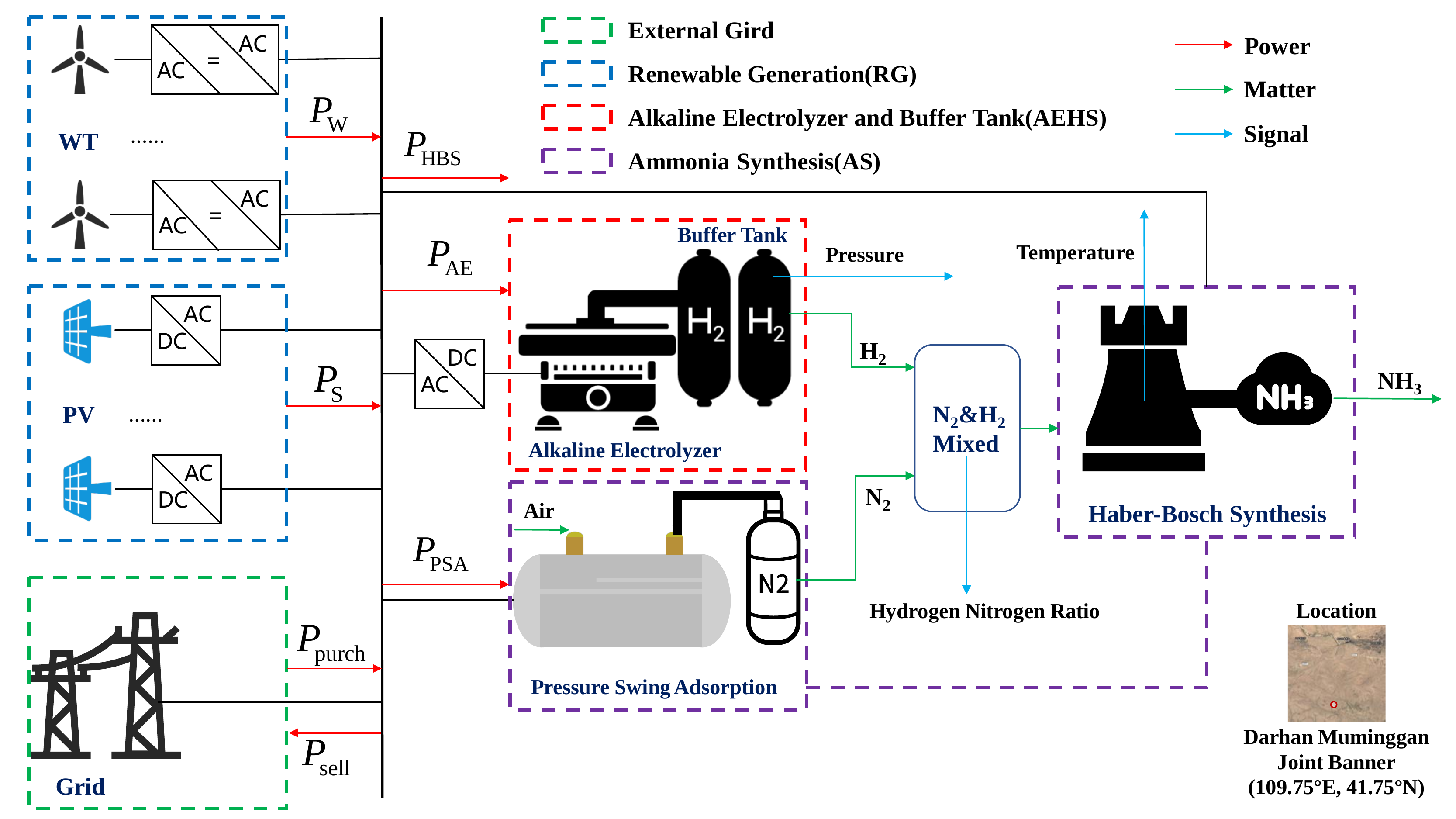}
  \caption{On-Grid renewable power to ammonia system integration and operation mode.}
  \label{fig:OperationTopu}
  \vspace{-12pt}
\end{figure}

\vspace{-12pt}
\section{RePtA System and Its Business Mode}
\label{sec:Motivation}
In this section, first, the topology of the RePtA system is introduced. Then, the business mode of such a multi-investor system is presented.

\vspace{-12pt}
\subsection{Components of RePtA Systems}
\label{sec:system_components}
\textcolor{black}{The proposed topology of the RePtA system is shown in Fig. \ref{fig:OperationTopu}, which is a real project located in Baotou, Inner Mongolia, China. The power side includes wind and solar power, as well as power exchange with the external grid (both on-grid and off-grid power). The load side consists of two parts. One is hydrogen production using alkaline electrolysis technology. The other is ammonia production, including nitrogen produced by pressure swing adsorption (PSA) and ammonia synthesis with Haber Bosch synthesis (HBS). The energy storage is mainly hydrogen storage using the buffer tank.}

\vspace{-12pt}
\subsection{Business Mode of RePtA Systems}
\label{sec:introduction_business}
\begin{figure}[t]
  \centering
  %\vspace{-12pt}
  \includegraphics[width=3.46in]{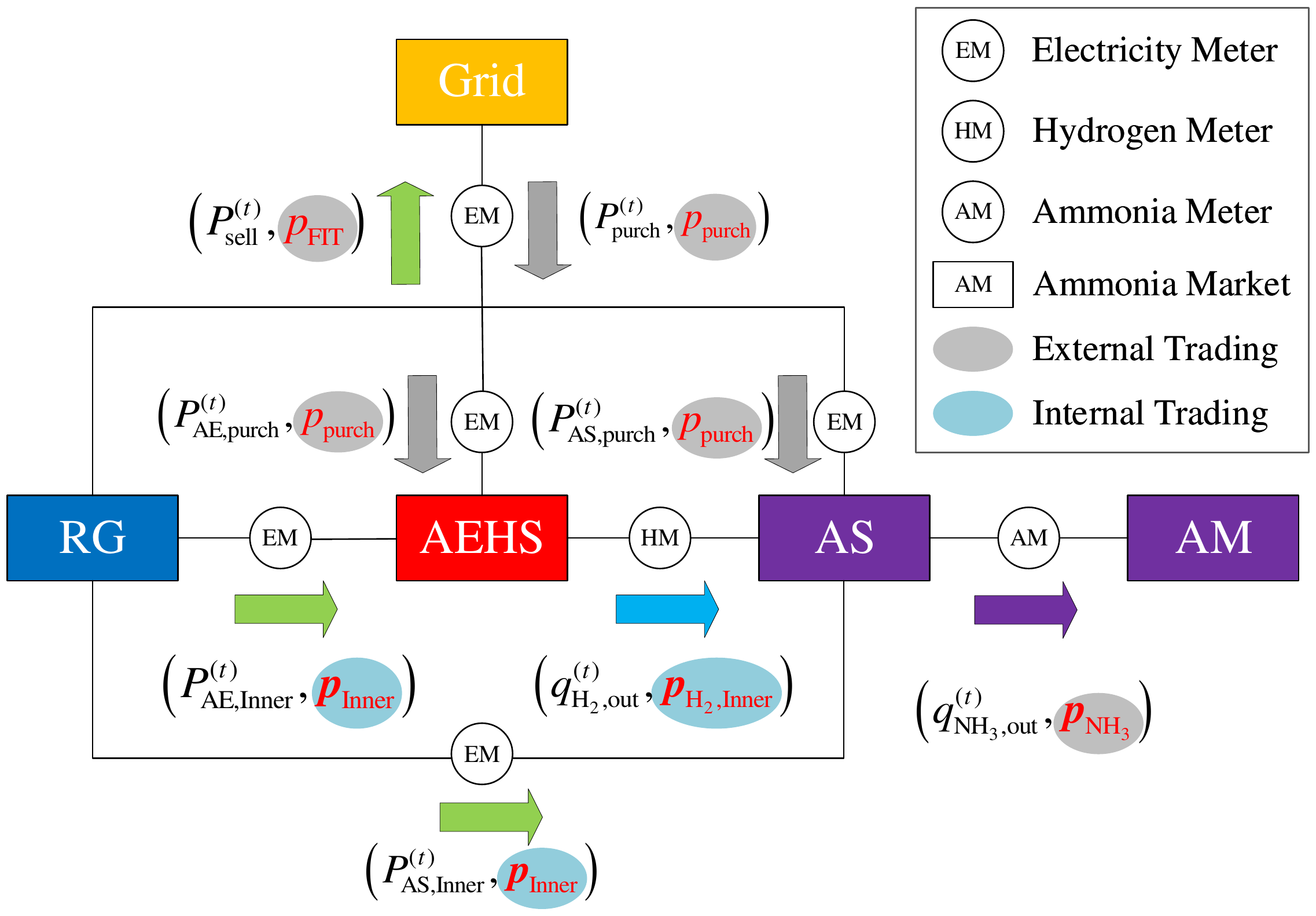}
  \caption{Business mode of RePtA including external and internal trading.}
  \label{fig:EconomicTopu}
  \vspace{-12pt}
\end{figure}

\textcolor{black}{To well discuss the business mode, the system is divided into three parts belonging to three different investors: the renewable energy generation (RG) part, the alkaline electrolyzer and hydrogen storage (AEHS) part, and the ammonia synthesis (AS) part, as shown in Fig. \ref{fig:EconomicTopu}. In such a multi-investor system, the trading mode is consist of two parts.}

\textcolor{black}{\emph{1) External trading in electricity and ammonia markets.} RG part sells the renewable energy to the grid, while AEHS and AS buy electricity from the grid respectively. And AS part sells the produced ammonia to the ammonia market, as the circle in gray plotted in Fig. \ref{fig:EconomicTopu}. Thus, the RePtA system plays the role of the price-taker in external trading, because the on/off grid electricity price and ammonia price are given by the markets.}

\textcolor{black}{\emph{2) Internal trading among RG, AEHS, and AS parts.} RG part sells the renewable energy to both AEHS and AS parts, and AEHS part sells the produced hydrogen to AS part, as the circle in blue plotted in Fig. \ref{fig:EconomicTopu}. Thus, the RePtA system plays the role of the price-maker in internal trading, because the inner electricity and hydrogen price are determined by the RG and AEHS parts respectively. }

\textcolor{black}{Therefore, planning, operation, and trading are deeply coupled in RePtA system design. And, limited flexibility model of ammonia synthesis and multi-investor economic model are established in Section \ref{sec:sa_model} and \ref{sec:MIE_model} respectively. }

%\vspace{-12pt}
\section{Limited Flexibility Model of Ammonia Synthesis and Model Verification}
\label{sec:sa_model}
In this section, first, limited flexibility model of ammonia synthesis is established. Then, actual dynamic regulation data of ammonia synthesis are used to identify the parameter in the proposed model, and the accuracy of the proposed model is analyzed.
\begin{figure}[t]
  \centering
  %\vspace{-12pt}
  \includegraphics[width=3.46in]{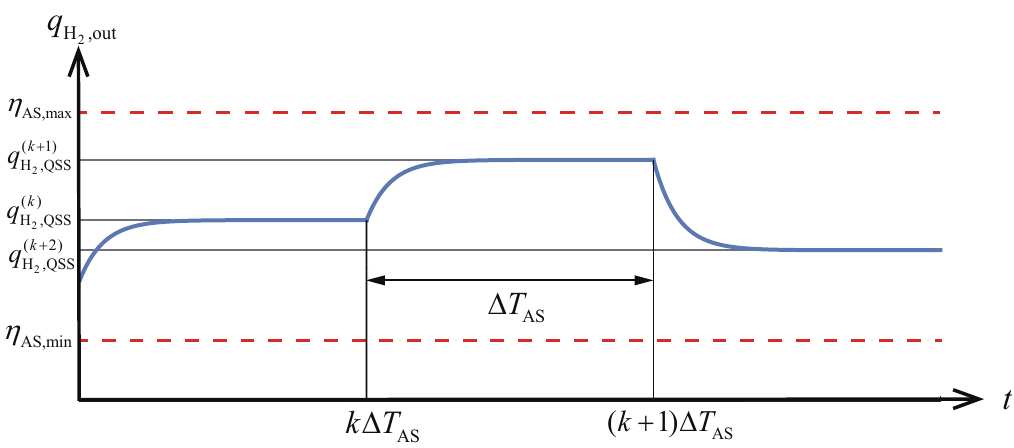}
  \caption{The proposed planned quasi-steady-state condition scheduling model for ammonia synthesis.}
  \label{fig:Ammonia_Adjustment_Fig}
  \vspace{-12pt}
\end{figure}

\begin{figure}[t]
  \centering
  %\vspace{-12pt}
  \includegraphics[width=3.4in]{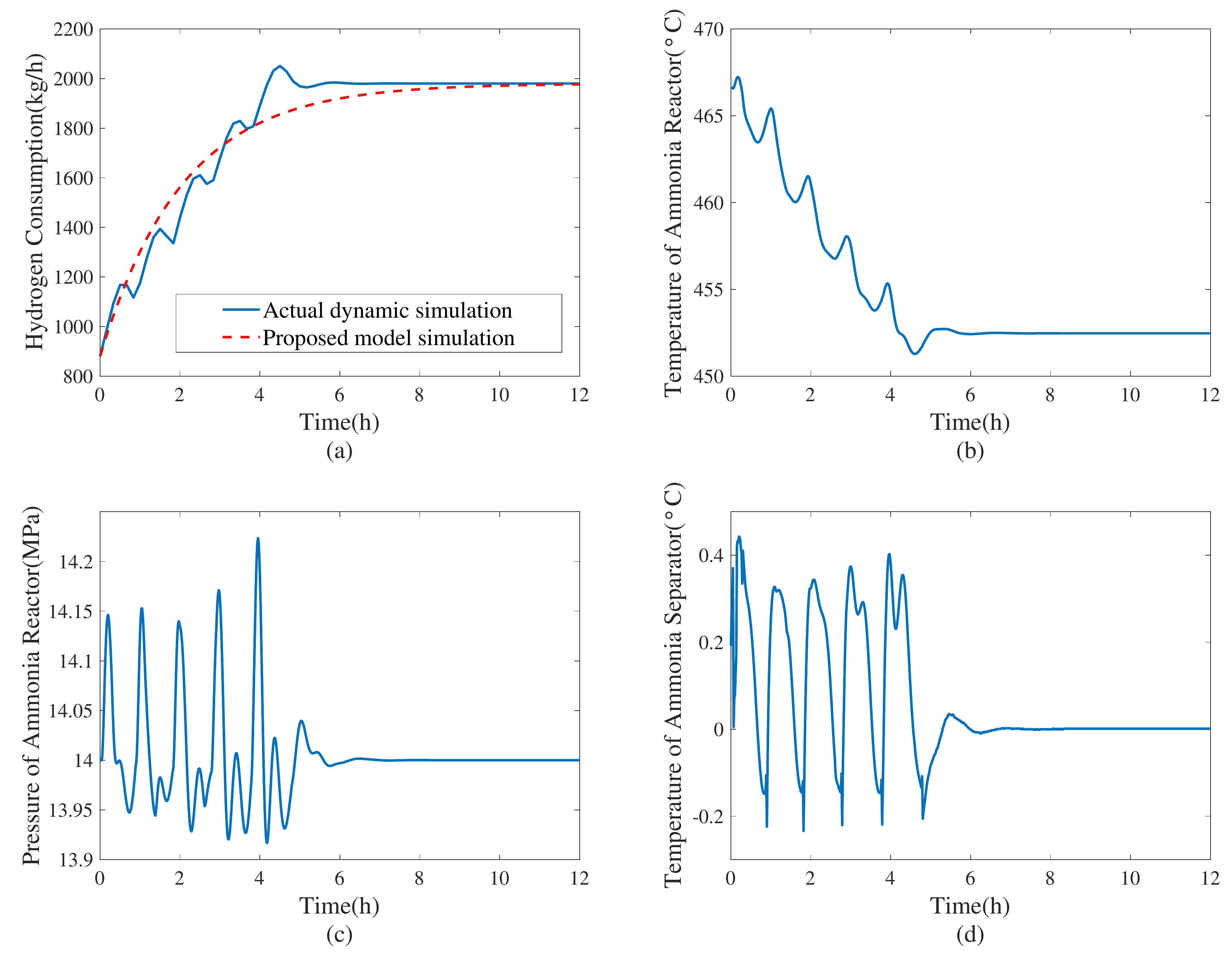}
  \caption{Dynamic regulation of ammonia synthesis under actual dynamic process and the identified proposed model. (a) Hydrogen consumption. (b) Temperature of Ammonia Reactor. (c) Pressure of Ammonia Reactor. (d) Temperature of Ammonia Separator.}
  \label{fig:Ammonia_Flexibility_Verify}
  \vspace{-12pt}
\end{figure}

\vspace{-12pt}
\subsection{Limited Flexibility Model of Ammonia Synthesis}
\label{sec:limit_sa_model}
The ammonia synthesis process converts hydrogen (H$_2$) and nitrogen (N$_2$) into ammonia (NH$_3$) and simultaneously consumes electrical energy.
The power consumption models of PSA and HBS are shown in (\ref{eq:psa}) and (\ref{eq:hbs}).
\begin{align}
   P_{{\rm{PSA}}}^{(t)} &= {\kappa _{{{\rm{N}}_{\rm{2}}}}}q_{{{\rm{N}}_{\rm{2}}}}^{(t)} \label{eq:psa} \\
   P_{{\rm{HBS}}}^{(t)} &= {\kappa _{{\rm{N}}{{\rm{H}}_{\rm{3}}}}}q_{{\rm{N}}{{\rm{H}}_3}}^{(t)} \label{eq:hbs}
\end{align}
\noindent

The values of $\kappa_{\rm{N_2}}$ and $\kappa_{\rm{NH_3}}$ can be found in \cite{armijo2020flexible,klyapovskiy2021optimal}, and a more detailed derivation process can be found in \cite{verleysen2020can}. According to the reaction equation of the ammonia synthesis process, the flow rate of each reaction material meets the following equation:
\begin{align}
   q_{{{\rm{H}}_{\rm{2}}}{\rm{,out}}}^{(t)}:q_{{{\rm{N}}_{\rm{2}}}}^{(t)}:q_{{\rm{N}}{{\rm{H}}_3}}^{(t)} = 3:1:2 \label{eq:reaction_eq}
\end{align}
\noindent

Combining (\ref{eq:psa})-(\ref{eq:reaction_eq}), the power consumption model of the ammonia synthesis process is expressed as
\begin{align}
   P_{{\rm{AS}}}^{(t)} = P_{{\rm{PSA}}}^{(t)}{\rm{ + }}P_{{\rm{HBS}}}^{(t)} = {\kappa _{{\rm{AS}}}}q_{{{\rm{H}}_{\rm{2}}}{\rm{,out}}}^{(t)}
\end{align}
\noindent
Where $q_{\mathrm{H_2,out}}^{(t)}$ is the hydrogen consumption rate, known as the working condition of ammonia synthesis. %The hydrogen production rate under the rated load is defined as the rated working condition of ammonia synthesis, denoted as
%\begin{align}
%   {q_{{{\rm{H}}_{\rm{2}}}{\rm{,r}}}} = \frac{{{C_{{\rm{AE}}}}}}{{{\kappa _{{{\rm{H}}_{\rm{2}}}}}}}
%\end{align}

The flexibility model of ammonia synthesis represented by the planned quasi-steady-state condition scheduling proposed in this paper, including condition scheduling (\ref{eq:qss_adjust}), load variation range (\ref{eq:sa_range}) and ramping limitations (\ref{eq:sa_ramping}), denoted as
\begin{align}
   &q_{{{\rm{H}}_{\rm{2}}}{\rm{,out}}}^{}(\tau ) = f\left( {q_{{{\rm{H}}_{\rm{2}}}{\rm{,QSS}}}^{(k)},q_{{{\rm{H}}_{\rm{2}}}{\rm{,QSS}}}^{(k + 1)};\tau } \right), \nonumber \\
   &\forall k \in \mathbb{K},\tau  \in \left[ {k\Delta {T_{{\rm{AS}}}},\left( {k + 1} \right)\Delta {T_{{\rm{AS}}}}} \right)\label{eq:qss_adjust}\\
   &{\eta _{{\rm{AS,min}}}}{q_{{{\rm{H}}_{\rm{2}}}{\rm{,r}}}} \le q_{{{\rm{H}}_{\rm{2}}}{\rm{,QSS}}}^{(k)} \le {\rm{ }}{\eta _{{\rm{AS,max}}}}{q_{{{\rm{H}}_{\rm{2}}}{\rm{,r}}}},\forall k \in \mathbb{K}\label{eq:sa_range} \\
   & - {r_ - }{q_{{{\rm{H}}_{\rm{2}}}{\rm{,r}}}} \le q_{{{\rm{H}}_{\rm{2}}}{\rm{,out}}}^{(t + 1)} - q_{{{\rm{H}}_{\rm{2}}}{\rm{,out}}}^{(t)} \le {r_ + }{q_{{{\rm{H}}_{\rm{2}}}{\rm{,r}}}},\forall t \in \mathbb{T} \label{eq:sa_ramping}
\end{align}
\noindent
where ${q_{{{\rm{H}}_{\rm{2}}}{\rm{,r}}}}$ is the rated working condition (nominal capacity) of ammonia synthesis. $q_{\rm{H_2,QSS}}^{(k)}$ is the $k$th quasi-steady-state condition of ammonia synthesis.

\textcolor{black}{$\Delta T_{\rm{AS}}$ is the scheduling period, shown in Fig. \ref{fig:Ammonia_Adjustment_Fig}, which is the key parameter to represent different levels of ammonia flexibility. For example, $\Delta T_{\rm{AS}} = 1 $ year means that ammonia synthesis is a fixed load without flexibility, while $\Delta T_{\rm{AS}} = 1 $ day denotes advanced flexibility of ammonia. This leads to multi-timescale operation issues, which must be modeled as a time series model instead of scenario-based model.}

For (\ref{eq:qss_adjust}), we use the common 1-order transition process in the chemical process as a reference:
\begin{align}
   q_{{{\rm{H}}_{\rm{2}}}{\rm{,out}}}^{}(\tau ) = q_{{{\rm{H}}_{\rm{2}}}{\rm{,QSS}}}^{(k + 1)} + \left( {q_{{{\rm{H}}_{\rm{2}}}{\rm{,QSS}}}^{(k)} - q_{{{\rm{H}}_{\rm{2}}}{\rm{,QSS}}}^{(k + 1)}} \right){e^{ - \frac{\tau }{{{T_{{\rm{trans}}}}}}}} \label{eq:sa_trans}
\end{align}
\noindent
where $T_{\rm{trans}}$ is the time constant of the transition process. \textcolor{black}{And, this model is demonstrated in the following section.}
\vspace{-12pt}
\subsection{Model Verification with Actual Dynamic Regulation Data}
\textcolor{black}{To further verify the proposed model, the actual dynamic regulation data of ammonia synthesis are presented in Fig. \ref{fig:Ammonia_Flexibility_Verify}, in which the rated load of ammonia synthesis $q_{\rm{H_2,r}} = 2200 \rm{kg}/h$, and load increases from $880 \rm{kg}/h$ ($40\% $load) to $1980 \rm{kg}/h$ ($90\%$load).}

\textcolor{black}{Using the real data to identify parameter in model (\ref{eq:sa_trans}), the result is that $T_{\rm{trans}}^{*}=2.066h$. The actual dynamic simulation and the identified model simulation are both plotted in Fig. \ref{fig:Ammonia_Flexibility_Verify} (a), and root mean square error (RMSE) between them is $63.9 \rm{kg}/h$ ($3\%$load). Furthermore, the temperature and pressure of ammonia reactor, as well as the temperature of ammonia separator are changed in a safety range\cite{rouwenhorst2019islanded}, shown in Fig. \ref{fig:Ammonia_Flexibility_Verify} (b)-(d).}

\textcolor{black}{Thus the proposed model guarantees accuracy, especially in hourly time resolution planning, and it is still a linear form without increasing the complexity in optimization.}

\vspace{-9pt}
\section{Multi-investor Economic Model of RePtA}
\label{sec:MIE_model}
In this section, first operation model and trading model for RePtA system are established repectively. Then the overall optimization model, i.e. multi-investor economic (MIE) model is formulated, with the discussion about the difficulties in solving and applying.

\vspace{-12pt}
\subsection{Operation Model for RePtA System}
\label{sec:operation_model}
Given the planning horizon $N$ and step length $\Delta T$, the set of time intervals $\mathbb{T}$ is defined as $\mathbb{T} = \left\{ 0,1,\ldots,N-1 \right\}$.

\subsubsection{Operation Model of Alkaline Electrolyzers}
\label{sec:AE_model}

Alkaline electrolyzers (AE) are used for converting power $P_{\mathrm{AE}}^{(t)}$ to hydrogen $q_{\mathrm{H_2,in}}^{(t)}$. At the planning level, a constant efficiency model can be adopted, as in  (\ref{eq:pAE}) \cite{li2021co}, and the variation load range is shown in (\ref{eq:cons_pAE}). %The ramping limitation is ignored in planning with an hourly time resolution due to the maximum ramping exceeding $\pm $ 20\%load/s \cite{armijo2020flexible}.
\vspace{-3pt}
\begin{gather}
   P_{{\rm{AE}}}^{(t)} = {\kappa _{{{\rm{H}}_{\rm{2}}}}}q_{{{\rm{H}}_{\rm{2}}}{\rm{,in}}}^{(t)} \label{eq:pAE} \\
   {\eta _{{\rm{AE,min}}}}{C_{{\rm{AE}}}} \le P_{{\rm{AE}}}^{(t)} \le {\eta _{{\rm{AE,max}}}}{C_{{\rm{AE}}}} \label{eq:cons_pAE} \\
   C_{\rm{AE}} = N_{\rm{AE}}C_{\rm{AE}}^{\rm{single}}
\end{gather}
\noindent
where ${\kappa _{{{\rm{H}}_{\rm{2}}}}}$ is the energy coverting coefficient of the alkaline electrolyzer. $C_{\mathrm{AE}}$ is the capacity of AE, $N_{\rm{AE}}$ is an integer variable representing the number of electrolyzers, and $C_{\rm{AE}}^{\rm{single}}$ is the single electrolyzer capacity, typical value is 5 MW.
%where ${\kappa _{{{\rm{H}}_{\rm{2}}}}}=\frac{{{\rm{LH}}{{\rm{V}}_{{{\rm{H}}_{\rm{2}}}}}}}{{{\eta _{{{\rm{H}}_{\rm{2}}}}}}}$ is the energy efficiency coefficient of the alkaline electrolyzer confirmed by the conversion efficiency ${\eta _{{{\rm{H}}_{\rm{2}}}}}$ and the lower heating value of hydrogen ${{\rm{LH}}{{\rm{V}}_{{{\rm{H}}_{\rm{2}}}}}}$. $C_{\mathrm{AE}}$ is the capacity of AE, $N_{\rm{AE}}$ is an integer variable representing the number of electrolyzers, and $C_{\rm{AE}}^{\rm{single}}$ is the capacity of a single electrolyzer; the typical value is 5 MW.

\subsubsection{Operation Model of Hydrogen Buffer Tanks}
\label{sec:HS_model}

The hydrogen buffer tank (HS) is used to balance the fluctuation caused by differences in flexibility between hydrogen and ammonia production, denoted as
\begin{gather}
   n_{{\rm{sto}}}^{(t + 1)} = n_{{\rm{sto}}}^{(t)} + \left( {q_{{{\rm{H}}_{\rm{2}}}{\rm{,in}}}^{(t)} - q_{{{\rm{H}}_{\rm{2}}}{\rm{,out}}}^{(t)}} \right)\Delta T,\ \forall t \in \mathbb{T} \label{eq:ssp_HS} \\
   {\eta _{{\rm{HS,min}}}}{C_{{\rm{HS}}}} \le n_{{\rm{sto}}}^{(t)} \le {\eta _{{\rm{HS,max}}}}{C_{{\rm{HS}}}},\ \forall t \in \mathbb{T} \label{eq:nsto_limit} \\
   n_{{\rm{sto}}}^{(0)} = n_{{\rm{sto}}}^{(N)} = 50\% {C_{{\rm{HS}}}}\label{eq:nsto_start_end}
\end{gather}
%\begin{align}
%   n_{{\rm{sto}}}^{(t + 1)} = n_{{\rm{sto}}}^{(t)} + \left( {q_{{{\rm{H}}_{\rm{2}}}{\rm{,in}}}^{(t)} - q_{{{\rm{H}}_{\rm{2}}}{\rm{,out}}}^{(t)}} \right)\Delta T,\ \forall t \in \mathbb{T}
%   \label{eq:ssp_HS}
%\end{align}
%\begin{align}
%   {\eta _{{\rm{HS,min}}}}{C_{{\rm{HS}}}} \le n_{{\rm{sto}}}^{(t)} \le {\eta _{{\rm{HS,max}}}}{C_{{\rm{HS}}}},\ \forall t \in \mathbb{T} \label{eq:nsto_limit}
%\end{align}
%\begin{align}
%   n_{{\rm{sto}}}^{(0)}{\rm{ = }}n_{{\rm{sto}}}^{(N)} = 50\% {C_{{\rm{HS}}}}\label{eq:nsto_start_end}
%\end{align}
\noindent
where $C_{\mathrm{HS}}$ is the capacity of the HS and $n_{\rm{sto}}^{(t)}$ is the hydrogen inventory in the tank. $q_{\rm{H_2,in}}^{(t)}$ is the flow-in rate related to hydrogen production by the electrolyzer, while $q_{\rm{H_2,out}}^{(t)}$ is the flow-out rate related to hydrogen consumption by ammonia synthesis. Considering the limitation of pressure in the buffer tank and to reserve some space for adjustment in dispatching and real-time control, (\ref{eq:nsto_limit}) is proposed. (\ref{eq:nsto_start_end}) means that hydrogen is completely consumed every year.

\subsubsection{Renewable Energy Generation}
\label{sec:renewable_generation}

If wind farms and PV plants have been built, the hourly historical data of wind power and solar power can be standardized as input ($P_{{\rm{W,sta}}}^{(t)}$, $P_{{\rm{S,sta}}}^{(t)}$). \textcolor{black}{The intermittency and volatility of renewable generation are reflected in these time series data.} Maximum output of wind and solar power can be described as (\ref{eq:pw}) and (\ref{eq:ps}) respectively, using the capacity of WT ($C_{\rm{W}}$) and PV ($C_{\rm{S}}$). If there is no historical data, it can be generated by using the local historical meteorological data; details can be found in \cite{deshmukh2008modeling}, which is not the focus of this paper.
%\cite{deshmukh2008modeling,hosseinalizadeh2016economic,klyapovskiy2021optimal,kolhe2003analytical}
\begin{align}
   %&\left\{ {P_{{\rm{W,norm}}}^{(t)},P_{{\rm{S,norm}}}^{(t)};t \in \mathbb{T}} \right\} \label{eq:rg_input}\\
   P_{\rm{W}}^{(t)} &= C_{\rm{W}}P_{\rm{W,sta}}^{(t)}\label{eq:pw}\\
   P_{\rm{S}}^{(t)} &= C_{\rm{S}}P_{\rm{S,sta}}^{(t)}\label{eq:ps}
\end{align}
%\begin{align}
%P_{\rm{W}}^{(t)} &= C_{\rm{W}}P_{\rm{W,norm}}^{(t)}\label{eq:pw}\\
%P_{\rm{S}}^{(t)} &= C_{\rm{S}}P_{\rm{S,norm}}^{(t)}\label{eq:ps}
%\end{align}

\subsubsection{Constraints of Power and Energy Exchanging with the Grid}
\label{sec:exchange_constraints}

From the perspective of power constraints, according to the measurement with a single meter, only power sales to the grid $P_{{\rm{sell}}}^{(t)}$ (on-grid) or power purchases from grid $P_{{\rm{purch}}}^{(t)}$ (off-grid) are allowed at the same time. The corresponding constraints are shown as follows:
\vspace{-3pt}
\begin{align}
   0 \le \ & P_{{\rm{sell}}}^{(t)} \le b_{{\rm{grid}}}^{(t)}{M_1} \\
   0 \le \ & P_{{\rm{purch}}}^{(t)} \le \left( {1 - b_{{\rm{grid}}}^{(t)}} \right){M_1}
\end{align}
\noindent
\vspace{-3pt}
where $b_{{\rm{grid}}}^{(t)}$ is a binary variable, $b_{{\rm{grid}}}^{(t)}=0$ represents off-grid power, and $b_{{\rm{grid}}}^{(t)}=1$ represents on-grid power. $M_1$ is a positive number that is sufficiently large.

From the perspective of energy constraints, a net on-grid energy constraint is adopted based on the real policy in Inner Mongolia\cite{MX-Ongrid-rate-2020}, denoted as
\begin{align}
   &\Delta T\sum\limits_{t \in \mathbb{T}}^{} {\left( {P_{{\rm{sell}}}^{(t)} - P_{{\rm{purch}}}^{(t)}} \right)}  \le {r_{\rm{net}}} \Delta T\sum\limits_{t \in \mathbb{T}}^{} {\left( {P_{{\rm{W}}}^{(t)} + P_{{\rm{S}}}^{(t)}} \right)} \label{eq:net_on_grid_cons}
\end{align}
\noindent
where $r_{\rm{net}}$ represents the maximum net on-grid rate given by the government. The left term of (\ref{eq:net_on_grid_cons}) is net on-grid energy, and the right term is the total annual renewable power generation.

%\begin{align}
%   {E_{{\rm{RG}}}} = \Delta T\sum\limits_{t \in \mathbb{T}}^{} {\left( {{C_{\rm{W}}}P_{{\rm{W,norm}}}^{(t)} + {C_{\rm{S}}}P_{{\rm{S,norm}}}^{(t)}} \right)} \label{eq:ERen}
%\end{align}
%\noindent
%where $C_{\mathrm{W}}$ and $C_{\mathrm{S}}$ are the capacities of WTs and PVs, respectively.

\subsubsection{Distribution Model of Power and Energy between Hydrogen and Ammonia}
\label{sec:distribution_model}
Since the power consumed by the AEHS and AS parts are mainly from renewable power supplied by the RG part, and partly from the external grid, to distinguish different power sources, four groups of continuous variables, namely, $P_{{\rm{AE,Inner}}}^{(t)}$, $P_{{\rm{AE,purch}}}^{(t)}$, $P_{{\rm{AS,Inner}}}^{(t)}$ and $P_{{\rm{AS,purch}}}^{(t)}$ are introduced, shown in Fig. \ref{fig:EconomicTopu}, with equality constraints:
\begin{align}
    &P_{{\rm{Inner}}}^{(t)} = P_{{\rm{AE,Inner}}}^{(t)} + P_{{\rm{AS,Inner}}}^{(t)} \nonumber \\
    &P_{{\rm{purch}}}^{(t)} = P_{{\rm{AE,purch}}}^{(t} + P_{{\rm{AS,purch}}}^{(t)}  \nonumber \\
    &P_{{\rm{AE}}}^{(t)} = P_{{\rm{AE,Inner}}}^{(t)} + P_{{\rm{AE,purch}}}^{(t)}  \nonumber \\
    &P_{{\rm{AS}}}^{(t)} = P_{{\rm{AS,Inner}}}^{(t)} + P_{{\rm{AS,purch}}}^{(t)},\ \forall t \in \mathbb{T} \label{eq:inner_distribution} \\
    &P_{{\rm{Inner}}}^{(t)} = P_{{\rm{W}}}^{(t)} + P_{{\rm{S}}}^{(t)} - P_{{\rm{sell}}}^{(t)} - P_{{\rm{curt}}}^{(t)} \label{eq:inner_power}
\end{align}
\noindent
where $P_{{\rm{Inner}}}^{(t)}$ is the power supplied by renewable power, and $P_{{\rm{curt}}}^{(t)}$ is the curtailment of renewable power.

\subsubsection{Operation Constraints Related to System Integration}
\label{sec:integration_cosntraints}

Two kinds of constraints need to be considered in system integration: one is the real-time power balance, and the other is the constraint of the annual output of ammonia, described by (\ref{eq:power_balance}) and (\ref{eq:ammonia_annual_output}), respectively.
%\begin{align}
%    &P_{{\rm{W}}}^{(t)} + P_{{\rm{S}}}^{(t)} + P_{{\rm{purch}}}^{(t)} \nonumber \\
%    &= P_{{\rm{sell}}}^{(t)} + P_{{\rm{curt}}}^{(t)} + P_{{\rm{AE}}}^{(t)} + P_{{\rm{AS}}}^{(t)},\ \forall t \in \mathbb{T} \label{eq:power_balance}
%\end{align}
\begin{gather}
    P_{{\rm{W}}}^{(t)} + P_{{\rm{S}}}^{(t)} + P_{{\rm{purch}}}^{(t)} = P_{{\rm{sell}}}^{(t)} + P_{{\rm{curt}}}^{(t)} + P_{{\rm{AE}}}^{(t)} + P_{{\rm{AS}}}^{(t)},\ \forall t \in \mathbb{T} \label{eq:power_balance}\\
    {m_{{\rm{N}}{{\rm{H}}_{\rm{3}}}}} = {{\rm{C}}_{{\rm{H2mA}}}}\Delta T\sum\limits_{t \in \mathbb{T}}^{} {q_{{{\rm{H}}_{\rm{2}}}{\rm{,out}}}^{(t)}} \leq {\bar{m}_{{\rm{N}}{{\rm{H}}_{\rm{3}}}}} \label{eq:ammonia_annual_output}
\end{gather}
\noindent
where ${{\rm{C}}_{{\rm{H2mA}}}}=5.060 \times 10^{-4}$ t/Nm$^3$ is the conversion factor from hydrogen to ammonia. $\bar{m}_{\rm{NH_3}}$ is nominal annual output of ammonia, while ${m}_{\rm{NH_3}}$ is real output.
\vspace{-12pt}
\subsection{Trading Model for RePtA System}
\label{sec:economic_model}

\subsubsection{RG Part}
\label{sec:RG}
In this part, the net annual revenue (\ref{eq:RGfunc}) is equal to annual incomes (\ref{eq:RGIncome}) minus the depreciation of the initial investment cost (\ref{eq:RGInvest}). The first term in (\ref{eq:RGIncome}) represents the income of renewable power sold to the grid under the feed-in-tariff (FIT) $p_{\mathrm{FIT}}$, while the second term represents that of the power sold to both the AEHS and AS parts under the inner electricity price $p_{\mathrm{Inner}}$.
\begin{align}
   {J_{{\rm{RG}}}} &= {J_{{\rm{RG,Profit}}}} - {J_{{\rm{RG,Invest}}}} \label{eq:RGfunc}\\
   {J_{{\text{RG,Profit}}}} &= \Delta T\sum\limits_{t \in \mathbb{T}}^{} {{p_{{\text{FIT}}}}P_{{\text{sell}}}^{(t)}}  + \Delta T\sum\limits_{t \in \mathbb{T}}^{} {{p_{{\text{Inner}}}}P_{{\text{Inner}}}^{(t)}} \label{eq:RGIncome} \\
   {J_{{\rm{RG,Invest}}}} &= {\rm{CRF}}(r,{Y_{{\rm{RG}}}}) {\left( {I_{\rm{S}}^{{\rm{init}}} + I_{\rm{S}}^{{\rm{O\& M}}}} \right){C_{\rm{S}}}}   \nonumber \\
   & {+ {\rm{CRF}}(r,{Y_{{\rm{RG}}}}) \left( {I_{\rm{W}}^{{\rm{init}}} + I_{\rm{W}}^{{\rm{O\& M}}}} \right){C_{\rm{W}}}}  \label{eq:RGInvest}
\end{align}
\noindent
%\begin{align}
%   {J_{{\rm{RG}}}} &= {J_{{\rm{RG,Invest}}}} - {J_{{\rm{RG,Profit}}}} \label{eq:RGfunc}\\
%   {J_{{\rm{RG,Invest}}}} &= {\rm{CRF}}(r,{Y_{{\rm{RG}}}}) {\left( {I_{\rm{S}}^{{\rm{init}}} + I_{\rm{S}}^{{\rm{O\& M}}}} \right){C_{\rm{S}}}}   \nonumber \\
%   & {+ {\rm{CRF}}(r,{Y_{{\rm{RG}}}}) \left( {I_{\rm{W}}^{{\rm{init}}} + I_{\rm{W}}^{{\rm{O\& M}}}} \right){C_{\rm{W}}}}  \label{eq:RGInvest}\\
%   {J_{{\text{RG,Profit}}}} &= \Delta T\sum\limits_{t \in \mathbb{T}}^{} {{p_{{\text{FIT}}}}P_{{\text{sell}}}^{(t)}}  + \Delta T\sum\limits_{t \in \mathbb{T}}^{} {{p_{{\text{Inner}}}}P_{{\text{Inner}}}^{(t)}} \label{eq:RGIncome}
%\end{align}
%\noindent
where ${\rm{CRF}}(r,Y) = \frac{{r{{\left( {1 + r} \right)}^Y}}}{{{{\left( {1 + r} \right)}^Y} - 1}}$ represents the capital recovery factor (CRF), $r = 8\%$\cite{li2021co} is the interest rate, and $Y$ denotes the lifetime of facilities. $I_{\rm{j}}^{{\rm{init}}}$ is the initial investment cost, and $I_{\rm{j}}^{{\rm{O\& M}}}$ denotes the operation and maintenance cost. $j$ can be any facility; e.g., $\rm{W}$ represents wind turbines. %$\rm{S}$ represents PV.

\subsubsection{AEHS Part}
\label{sec:AEHS}

In this part, the net annual revenue (\ref{eq:AEHSfunc}) is equal to the annual incomes (\ref{eq:AEHSIncome}) minus the depreciation of the initial investment cost (\ref{eq:AEHSInvest}). The first term in (\ref{eq:AEHSIncome}) represents the income of hydrogen sold to the AS part with inner hydrogen price $p_{\mathrm{H_2,Inner}}$, while the second term represents the cost of purchasing electricity under inner electricity price $p_{\mathrm{Inner}}$ from the RG part and electricity price $p_{\mathrm{purch}}$ from the power grid.
\begin{align}
   &{J_{{\rm{AEHS}}}} = {J_{{\rm{AEHS,Profit}}}} - {J_{{\rm{AEHS,Invest}}}}  \label{eq:AEHSfunc}\\
   &{J_{{\rm{AEHS,Profit}}}} = \Delta T\sum\limits_{t \in \mathbb{T}}^{} {{p_{{{\rm{H}}_{\rm{2}}}{\rm{,Inner}}}}q_{{{\rm{H}}_{\rm{2}}}{\rm{,in}}}^{(t)}}   \nonumber \\
    &- \Delta T\sum\limits_{t \in \mathbb{T}}^{} {\left( {{p_{{\rm{Inner}}}}P_{{\rm{AE,Inner}}}^{(t)} + {p_{{\rm{purch}}}}P_{{\rm{AE,purch}}}^{(t)}}\label{eq:AEHSIncome} \right)}\\
    &{J_{{\rm{AEHS,Invest}}}} = {\rm{CRF}}(r,{Y_{{\rm{AEHS}}}})\left[ {\left( {I_{\rm{AE}}^{{\rm{init}}} + I_{\rm{AE}}^{{\rm{O\& M}}}} \right){C_{\rm{AE}}}} \right. \nonumber \\
   &\left. { + \left( {I_{\rm{HS}}^{{\rm{init}}} + I_{\rm{HS}}^{{\rm{O\& M}}}} \right){C_{\rm{HS}}}} \right] \label{eq:AEHSInvest}
\end{align}
\noindent

\subsubsection{AS Part}
\label{sec:AS}

In this part, the net annual revenue (\ref{eq:ASfunc}) is equal to the annual incomes (\ref{eq:sa_profit}) minus the depreciation of the initial investment cost (\ref{eq:ASInvest}). The first term in (\ref{eq:sa_profit}) represents the income of ammonia sold to the ammonia market (AM) under ammonia price $p_{\mathrm{NH_3}}$, and the second term means the cost of purchasing electricity, while the third term is the cost of purchasing hydrogen from AEHS.
\begin{align}
   &{J_{{\rm{AS}}}} = {J_{{\rm{AS,Profit}}}} - {J_{{\rm{AS,Invest}}}} \label{eq:ASfunc}\\
   &{J_{{\rm{AS,Profit}}}} = \Delta T\sum\limits_{t \in \mathbb{T}}^{} {{p_{{\rm{N}}{{\rm{H}}_{\rm{3}}}}}{{\text{C}}_{{\rm{H2mA}}}}q_{{{\rm{H}}_{\rm{2}}}{\rm{,out}}}^{(t)}}  \nonumber \\
    &- \Delta T\sum\limits_{t \in \mathbb{T}}^{} {\left( {{p_{{\rm{Inner}}}}P_{{\rm{AS,Inner}}}^{(t)} + {p_{{\rm{purch}}}}P_{{\rm{AS,purch}}}^{(t)}} \right)}   \nonumber \\
    &- \Delta T\sum\limits_{t \in \mathbb{T}}^{} {{p_{{{\rm{H}}_{\rm{2}}}{\rm{,Inner}}}}q_{{{\rm{H}}_{\rm{2}}}{\rm{,out}}}^{(t)}} \label{eq:sa_profit}\\
    &{J_{{\rm{AS,Invest}}}} = {\rm{CRF}}(r,{Y_{{\rm{AS}}}})\left[ {\left( {I_{\rm{AS}}^{{\rm{init}}} + I_{\rm{AS}}^{{\rm{O\& M}}}} \right)} \right] \label{eq:ASInvest}
\end{align}

From the perspective of optimal economic benefits, the net revenue of the RG, AEHS and AS parts are summed as the objective function, denoted as
\vspace{-3pt}
\begin{align}
   {\rm{ }}J = {J_{{\rm{RG}}}} + {J_{{\rm{AEHS}}}} + {J_{{\rm{AS}}}} \label{eq:overall_obj}
\end{align}
\noindent
\vspace{-3pt}
\vspace{-12pt}
\subsubsection{Earnings Ratio Constraints of Multi-Investors}
\label{sec:ER_cons}
Different investors generally have requirements for the minimum earnings ratio, corresponding constraints can be expressed as
\begin{align}
    &{\rm{E}}{{\rm{R}}_{{\rm{RG,min}}}}{J_{{\rm{RG,Invest}}}} \leqslant {J_{{\rm{RG}}}} \nonumber \\
    &{\rm{E}}{{\rm{R}}_{{\rm{AEHS,min}}}}{J_{{\rm{AEHS,Invest}}}} \leqslant {J_{{\rm{AEHS}}}} \nonumber \\
    &{\rm{E}}{{\rm{R}}_{{\rm{AS,min}}}}{J_{{\rm{AS,Invest}}}} \leqslant {J_{{\rm{AS}}}} \label{eq:cons_ER}
\end{align}
\noindent
where $\rm{ER}_{\rm{RG,min}}$, $\rm{ER}_{\rm{AEHS,min}}$ and $\rm{ER}_{\rm{AS,min}}$ are minimum earnings ratios for different investors and generally set to 0.

\vspace{-12pt}
\subsection{Multi-Investor Economic Model and Analysis}
\label{sec:overall_model}

\subsubsection{Multi-Investor Economic Model Formulation}
\label{sec:overall_model_formulation}

Summarizing all the above, the overall optimization model, i.e. Multi-Investor economic (MIE) model is established, denoted as
\begin{align}
   \begin{array}{l}
   \mathop {\max }\limits_{\bm{U}} \left( {\ref{eq:overall_obj}} \right)\\
   {\rm{s}}{\rm{.t}}{\rm{. }}\left( \ref{eq:psa} \right) - \left( \ref{eq:ASInvest} \right),\left( {\ref{eq:cons_ER}} \right)
   \end{array}
   \label{eq:overall_opt}
\end{align}
\noindent
where ${\bm{U}} = \left\{ {{C_{\rm{S}}},{C_{\rm{W}}},{C_{{\rm{AE}}}},{C_{{\rm{HS}}}},{p_{{\rm{Inner}}}},{p_{{{\rm{H}}_{\rm{2}}}{\rm{,Inner}}}}} \right\}$ represents the decision variables \textcolor{black}{related to the system design}, including the capacity of the system components, inner electricity and hydrogen price. \textcolor{black}{And system operations are also decision variables, which are not listed in $\bm{U}$, shown in Fig. \ref{fig:Algorithm_FlowChart}.}

\subsubsection{Difficulty in Solving}
\label{sec:solving_difficulty}
The nonconvexity of the optimization model (\ref{eq:overall_opt}) is very obvious. There are mainly two kinds of bilinear terms in the model. One is the coupling between design parameters, such as $p_{\rm{Inner}}C_{\rm{W}}$ and $p_{\rm{Inner}}C_{\rm{S}}$. The number of such bilinear terms is limited and can be well treated by some convex relaxation methods. The other is the coupling of design parameters and operation variables, such as $p_{\rm{Inner}}P_{\rm{sell}}^{(t)}$, $p_{\rm{Inner}}P_{\rm{AE,Inner}}^{(t)}$, and $p_{\rm{H_2,Inner}}q_{\rm{H_2,in}}^{(t)}$. Compared with the former, this kind of bilinear term is too large to be handled without increasing the complexity. This indicates that planning, operation and trading of RePtA system are deeply coupled, leading to the difficulty in solving.

In other words, large-scale mixed-integer nonlinear programming (MINLP) problems are difficult to solve accurately and efficiently at present.

\vspace{-12pt}
\section{Two-Stage Decomposed Sizing and Pricing Method for Solving MIE Model}
\label{sec:two_stage_model}

In this section, to address the difficulties in solving and applying the MIE model (\ref{eq:overall_opt}), a two-stage decomposed sizing and pricing method is proposed. First, a proposition is introduced with proof, which indicates that sizing and pricing in order is equivalent to the MIE model (\ref{eq:overall_opt}). Second, the sizing optimization model is proposed to determine the capacity of all facilities, and a robust model based on information gap decision theory (IGDT) is proposed, to handle the uncertainty of renewable generation. Then, the pricing optimization model is proposed to balance the earning ratio of each investor. Finally, the framework of the proposed method is given, which is convenient for applying in practical engineering.

\vspace{-12pt}
\subsection{Proposition with Proof}
\label{sec:proof_prop}
\textcolor{black}{We have the following proposition:}

\textcolor{black}{\emph{Proposition 1:} The objective function of the proposed MIE model (\ref{eq:overall_opt}), i.e., $J$, is independent of inner electricity and hydrogen prices ($p_{\mathrm{Inner}}$ and $p_{\mathrm{H_2,Inner}}$).}

See the proof as follows.

In Section \ref{sec:economic_model}, the objective function $J$ is defined in (\ref{eq:RGfunc})-(\ref{eq:overall_obj}), and the part related to ${p_{{\rm{Inner}}}}$ and ${p_{{{\rm{H}}_{\rm{2}}}{\rm{,Inner}}}}$ in the objective function $J$, is written as $J'$, denoted as
\begin{align}
   J' = \  &\Delta T\sum\limits_{t \in \mathbb{T}}^{} {{p_{{\rm{Inner}}}} P_{{\rm{Inner}}}^{(t)} } - \Delta T\sum\limits_{t \in \mathbb{T}}^{} { {{p_{{\rm{Inner}}}}P_{{\rm{AE,Inner}}}^{(t)}}} \nonumber \\
   &  - \Delta T\sum\limits_{t \in \mathbb{T}}^{} { {{p_{{\rm{Inner}}}}P_{{\rm{AS,Inner}}}^{(t)}}}+ \Delta T\sum\limits_{t \in \mathbb{T}}^{} {{p_{{{\rm{H}}_{\rm{2}}}{\rm{,Inner}}}}q_{{{\rm{H}}_{\rm{2}}}{\rm{,in}}}^{(t)}} \nonumber \\
   &  - \Delta T\sum\limits_{t \in \mathbb{T}}^{} {{p_{{{\rm{H}}_{\rm{2}}}{\rm{,Inner}}}}q_{{{\rm{H}}_{\rm{2}}}{\rm{,out}}}^{(t)}}
\end{align}
and then calculated as
\begin{align}
    J' = \ &{p_{{\rm{Inner}}}}\Delta T\sum\limits_{t \in \mathbb{T}}^{} {\left( {P_{{\rm{Inner}}}^{(t)} - P_{{\rm{AE,Inner}}}^{(t)} - P_{{\rm{AS,Inner}}}^{(t)}} \right)} \nonumber \\
    &+ {p_{{{\rm{H}}_{\rm{2}}}{\rm{,Inner}}}}\Delta T\sum\limits_{t \in \mathbb{T}}^{} {\left( {q_{{{\rm{H}}_{\rm{2}}}{\rm{,in}}}^{(t)} - q_{{{\rm{H}}_{\rm{2}}}{\rm{,out}}}^{(t)}} \right)}
    \label{eq:J_inner_func}
\end{align}
According to (\ref{eq:inner_distribution}), the first term on the right-hand side of (\ref{eq:J_inner_func}) is equal to 0. From (\ref{eq:ssp_HS}), we have
\begin{align}
    \sum\limits_{t \in \mathbb{T}}^{} {n_{{\rm{sto}}}^{(t + 1)}}  = \sum\limits_{t \in \mathbb{T}}^{} {n_{{\rm{sto}}}^{(t)}}  + \sum\limits_{t \in \mathbb{T}}^{} {\left( {q_{{{\rm{H}}_{\rm{2}}}{\rm{,in}}}^{(t)} - q_{{{\rm{H}}_{\rm{2}}}{\rm{,out}}}^{(t)}} \right)\Delta T}
\end{align}
which is further calculated as
\begin{align}
    \sum\limits_{t \in \mathbb{T}}^{} {\left( {q_{{{\rm{H}}_{\rm{2}}}{\rm{,in}}}^{(t)} - q_{{{\rm{H}}_{\rm{2}}}{\rm{,out}}}^{(t)}} \right)\Delta T}  = n_{{\rm{sto}}}^{(N)} - n_{{\rm{sto}}}^{(0)} = 0
\end{align}
Therefore, the second term on the right-hand side of (\ref{eq:J_inner_func}) is also equal to 0, i.e.
\begin{align}
    J' = 0 \label{eq:proof_result}
\end{align}
Finally, (\ref{eq:proof_result}) concludes this proof.

\textcolor{black}{Since the \emph{Proposition 1} is proved,  we reveal that the sizing subproblem and the pricing subproblem can be decoupled in order. Thus, the sizing model and the pricing model are introduced in the following sections respectively.}

%\vspace{-12pt}
\subsection{Stage I: The Sizing Optimization Model}
\label{sec:sizing_model}

\subsubsection{Deterministic Sizing Model}
\label{sec:D_sizing}

The sizing model in stage I does not consider the constraints of the multi-investor economic, and only considers the maximization of the overall net revenue of the whole system, denoted as
\begin{align}
   \begin{array}{l}
   DTR \equiv \mathop {\max }\limits_{\bm{U'}} \left( {\ref{eq:overall_obj}} \right)\\
   {\rm{s}}{\rm{.t}}{\rm{. }}\left( \ref{eq:psa} \right) - \left( \ref{eq:net_on_grid_cons} \right),\left( {\ref{eq:power_balance}} \right) - \left( {\ref{eq:ASInvest}} \right)
   \end{array}
   \label{eq:sizing_opt}
\end{align}
\noindent
where ${\bm{U'}} = \left\{ {{C_{\rm{S}}},{C_{\rm{W}}},{C_{{\rm{AE}}}},{C_{{\rm{HS}}}}} \right\}$ denotes the decision variables related to system design, including only the capacity of the system components. Other variables related to operation are shown in Fig. \ref{fig:Algorithm_FlowChart}. $DTR$ is the deterministic total revenue

Compared to the overall optimization model (\ref{eq:overall_opt}), the sizing model (\ref{eq:sizing_opt}) ignores the earnings ratio constraints of multiple investors in (\ref{eq:cons_ER}). Thus, there is no need to consider the power distribution between hydrogen and ammonia production in (\ref{eq:inner_distribution}). Model (\ref{eq:sizing_opt}) is also a typical MILP problem, which can be solved using commercial solvers such as \emph{Gurobi}.

%\vspace{-12pt}
\subsubsection{IGDT-Based Robust Sizing Model}
\textcolor{black}{To handle the uncertainty of renewable generation, a robust model based on IGDT \cite{7994723,daneshvar2020novel,7346507} is proposed. The uncertainty of wind and solar power are described as the robust region $\pmb{\mathcal{U}}$, determined by the uncertainty horizon ($\alpha$) and forecasted/estimated renewable power ($\hat{P}_{\rm{W}}^{(t)}$ and  $\hat{P}_{\rm{S}}^{(t)}$), denoted as}
\begin{align}
   P_{\rm{W}}^{(t)} \in \pmb{\mathcal{U}} \left( \alpha,\hat{P}_{\rm{W}}^{(t)} \right) =  \left\{ P_{\rm{W}}^{(t)} \left| {\left | {\frac{ P_{\rm{W}}^{(t)} - \hat{P}_{\rm{W}}^{(t)} }{\hat{P}_{\rm{W}}^{(t)}} } \right |} \leq \alpha \right. \right\} \label{eq:uncertainty_pw} \\
   P_{\rm{S}}^{(t)} \in \pmb{\mathcal{U}} \left( \alpha,\hat{P}_{\rm{S}}^{(t)} \right) =  \left\{ P_{\rm{S}}^{(t)} \left| {\left | {\frac{ P_{\rm{S}}^{(t)} - \hat{P}_{\rm{S}}^{(t)} }{\hat{P}_{\rm{S}}^{(t)}} } \right |} \leq \alpha \right. \right\} \label{eq:uncertainty_ps}
\end{align}

\textcolor{black}{IGDT-based robust model is to maximize the uncertainty horizon in (\ref{eq:IGDT_1}), while robust total revenue ($RTR$) is bounded by means of $\beta$ and $DTR$ in (\ref{eq:IGDT_2}), under the worst-case, i.e. uncertainty variables take their lower bounds as given in (\ref{eq:IGDT_3}).}
\vspace{-12pt}
\begin{align}
   %\begin{array}{l}
   \mathop {\max }\limits_{\bm{U'}} \; & \alpha  \label{eq:IGDT_1} \\
   {\rm{s}}{\rm{.t}}{\rm{. }} \; &RTR = J \geq (1-\beta)DTR  \label{eq:IGDT_2} \\
   &P_{\rm{W}}^{(t)} = (1-\alpha){\hat{P}}_{\rm{W}}^{(t)}, P_{\rm{S}}^{(t)} = (1-\alpha){\hat{P}}_{\rm{S}}^{(t)} \label{eq:IGDT_3}  \\
   &C_{\rm{W}} = C_{\rm{W}}^{*}, C_{\rm{S}} = C_{\rm{S}}^{*},C_{\rm{AE}} = C_{\rm{AE}}^{*}  \label{eq:IGDT_4} \\
   &\left( \ref{eq:psa} \right) - \left( \ref{eq:net_on_grid_cons} \right),\left( {\ref{eq:power_balance}} \right) - \left( {\ref{eq:ASInvest}} \right) \label{eq:IGDT_5}
   %\end{array}
\end{align}
\vspace{-12pt}

\vspace{-12pt}
\subsection{Stage II: The Pricing Optimization Model}
\label{sec:pricing_model}
%\textcolor{red}{This part is doubted by reviewers, the objective is not reflecting the maximizing the revenue of each investor?}
According to the results of stage I, \textcolor{black}{all investors have possibility to get the maximum revenue because the maximum revenue of the whole system has been guaranteed in stage I. Thus, the objective of the pricing model in stage II is to balance the interest demands of different investors, i.e. minimizing the earnings ratio deviation of each investor as an objective function, denoted as}
\begin{align}
    J_2 = \left| {E{R_{{\rm{RG}}}} - E{R_{{\rm{AEHS}}}}} \right| + \left| {E{R_{{\rm{AEHS}}}} - E{R_{{\rm{AS}}}}} \right|
\end{align}
\noindent
where the earnings ratio of each investor can be expressed in a linear form using the solution of model (\ref{eq:sizing_opt}):
\begin{align}
   &E{R_{{\rm{RG}}}} = \left( {{J_{{\rm{RG,Profit}}}} - J_{{\rm{RG,Invest}}}^*} \right)/J_{{\rm{RG,Invest}}}^* \nonumber \\
   &E{R_{{\rm{AEHS}}}} = \left( {{J_{{\rm{AEHS,Profit}}}} - J_{{\rm{AEHS,Invest}}}^*} \right)/J_{{\rm{AEHS,Invest}}}  \nonumber \\
   &E{R_{{\rm{AS}}}} = \left( {{J_{{\rm{AS,Profit}}}} - J_{{\rm{AS,Invest}}}^*} \right)/J_{{\rm{AS,Invest}}}^*
   \label{eq:er_2}
\end{align}

%(\ref{eq:inner_distribution}) and (\ref{eq:cons_ER}) can be further denoted as
%\begin{align}
%   &E{R_{{\rm{RG,min}}}} \le E{R_{{\rm{RG}}}}  \nonumber \\
%   &E{R_{{\rm{AEHS,min}}}} \le E{R_{{\rm{AEHS}}}}   \nonumber \\
%   &E{R_{{\rm{AS,min}}}} \le E{R_{{\rm{AS}}}}
%\end{align}

%\begin{align}
%   &E{R_{{\rm{RG,min}}}} \le E{R_{{\rm{RG}}}} \le E{R_{{\rm{RG,max}}}} \nonumber \\
%   &E{R_{{\rm{AEHS,min}}}} \le E{R_{{\rm{AEHS}}}} \le E{R_{{\rm{AEHS,max}}}}  \nonumber \\
%   &E{R_{{\rm{AS,min}}}} \le E{R_{{\rm{AS}}}} \le E{R_{{\rm{AS,max}}}}
%\end{align}

Based on the optimal solution in stage I, most bilinear terms described in Section \ref{sec:solving_difficulty} are reduced to linear terms, and only bilinear terms ${p_{{\rm{Inner}}}}P_{{\rm{AE,Inner}}}^{(t)}$ and ${p_{{\rm{Inner}}}}P_{{\rm{AS,Inner}}}^{(t)}$ remain. By introducing variables ${E_{{\rm{AE,Inner}}}}$ and ${E_{{\rm{AS,Inner}}}}$ and equality constraints (\ref{eq:eInner}), the bilinear terms that need to be processed are greatly reduced.
%\vspace{-6pt}
\begin{align}
   {E_{{\rm{AE,Inner}}}} &= \Delta T \sum\limits_{t \in \mathbb{T}}^{} {P_{{\rm{AE,Inner}}}^{(t)}} \nonumber \\
   {E_{{\rm{AS,Inner}}}} &= \Delta T \sum\limits_{t \in \mathbb{T}}^{} {P_{{\rm{AS,Inner}}}^{(t)}} \label{eq:eInner}
\end{align}
%\vspace{-6pt}
%To address the bilinear terms ${p_{{\rm{Inner}}}}{E_{{\rm{AE,Inner}}}}$ and ${p_{{\rm{Inner}}}}{E_{{\rm{AS,Inner}}}}$, ${p_{{\rm{Inner}}}}$ can be discretized, and the big-M method is used to handle the bilinear terms as mixed integer linear terms, more details can be found in \cite{li2018operation}.

To well address the bilinear terms ${p_{{\rm{Inner}}}}{E_{{\rm{AE,Inner}}}}$ and ${p_{{\rm{Inner}}}}{E_{{\rm{AS,Inner}}}}$, ${p_{{\rm{Inner}}}}$ is discretized as
\begin{align}
   p_{\rm{Inner}} = \underline{p} + \frac{{\left( {\overline{p} - \underline{p}} \right)}}{{{N_p}}}\sum\limits_{j = 0}^{{N_p} - 1} {{b_j}} \label{eq:price_discrete}
\end{align}
\noindent
where $\underline{p}$ and $\overline{p}$ are the lower and upper boundaries of $p_{\rm{Inner}}$, respectively, and $b_j$ is a binary variable. Thus, the bilinear term ${p_{{\rm{Inner}}}} \omega$ is replaced with the linear combination of ${b_j} \omega$($\omega  \in \left\{ {{E_{{\rm{AE,Inner}}}},{E_{{\rm{AS,Inner}}}}} \right\}$). Then, the big-M method is used to address ${b_j} \omega$ as mixed integer linear terms; more details can be found in \cite{li2018operation}.

The objective function containing the 1-norm can be processed into linear form in the following ways. Thus, the optimization problem in Stage II can be expressed as follows:
\begin{align}
    \mathop {\min }\limits_{\bm{U''}} \ &{w_1} + {w_2} \nonumber \\
    {\rm{s}}{\rm{.t}}{\rm{. }} &- {w_1} \le E{R_{{\rm{RG}}}} - E{R_{{\rm{AEHS}}}} \le {w_1} \nonumber \\
   & - {w_2} \le E{R_{{\rm{AEHS}}}} - E{R_{{\rm{AS}}}} \le {w_2} \nonumber \\
    & \left( {\ref{eq:inner_distribution}} \right), \left( {\ref{eq:inner_power}} \right), \left( {\ref{eq:cons_ER}} \right), \left( {\ref{eq:er_2}} \right) - \left( {\ref{eq:eInner}} \right) \nonumber \\
    & Solution \  of  \   Model \left( \ref{eq:sizing_opt} \right)
    \label{eq:pricing_opt}
\end{align}
\noindent
where $\bm{U''} = \left\{ {{p_{{\rm{Inner}}}},{p_{{{\rm{H}}_{\rm{2}}}{\rm{,Inner}}}}} \right\}$ denotes the decision variables related to system design in stage II, including the inner electricity and hydrogen prices. (\ref{eq:pricing_opt}) is the mixed-integer linear programming (MILP), which can also be solved using \emph{Gurobi}.
\begin{figure}[t]
  \centering
  %\vspace{-12pt}
  \includegraphics[width=3.0in]{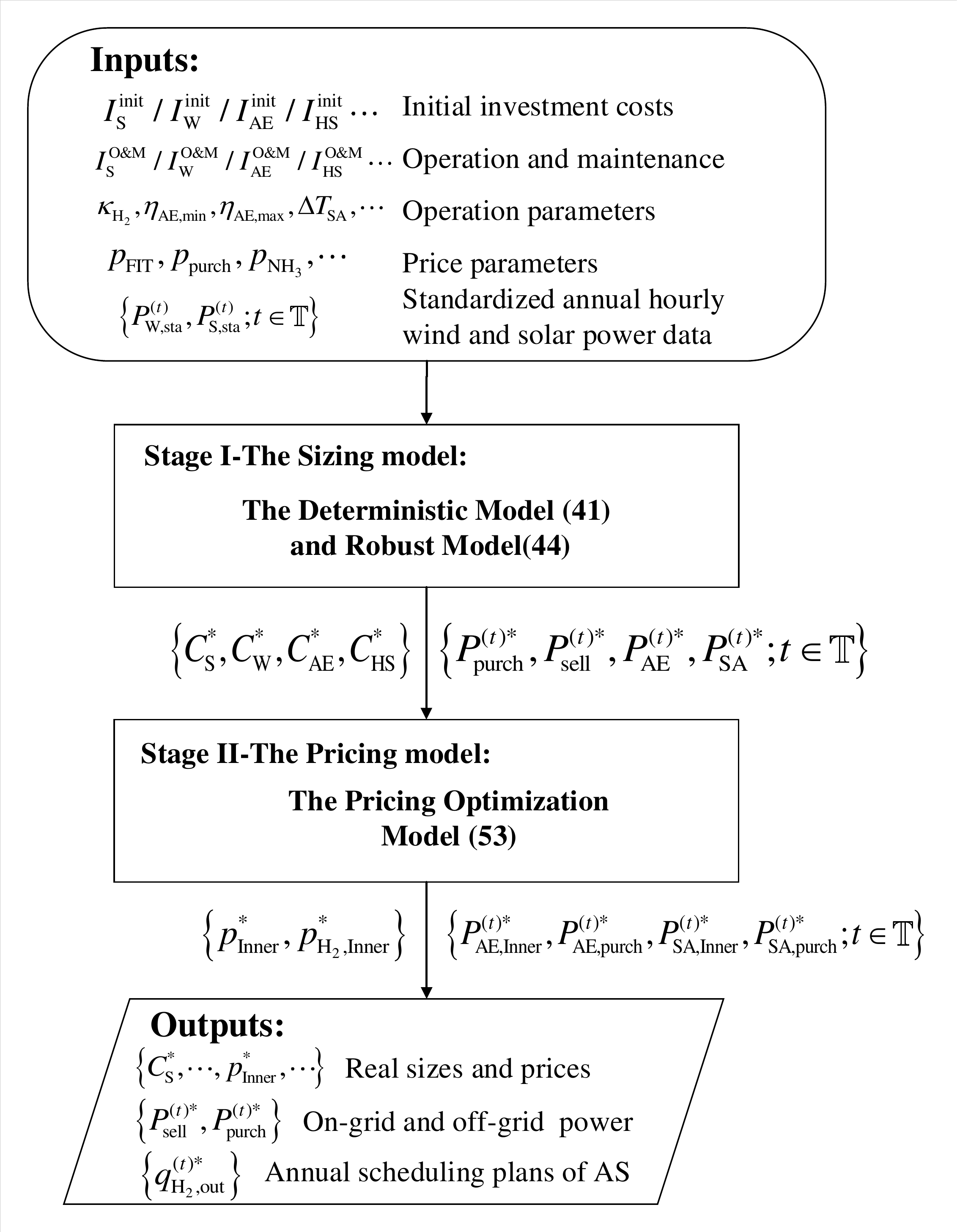}
  \caption{Framework of two-stage decomposed sizing and pricing method.}
  \label{fig:Algorithm_FlowChart}
  \vspace{-12pt}
\end{figure}

\vspace{-12pt}
\subsection{The Solution and Application of the Proposed Method}
\label{sec:model_application}
\textcolor{black}{The framework of the proposed two-stage decomposed sizing and pricing method is shown in Fig. \ref{fig:Algorithm_FlowChart}. In the application of practical engineering, boundary conditions, such as hourly wind and solar power curves, investment parameters, operation parameters, and price parameters, are used as input. Then the sizing model (\ref{eq:sizing_opt}) and the IGDT-based model (\ref{eq:pricing_opt}) are solved in turn, to obtain the robust solution of system capacities and operation mode, including the on/off grid power and annual scheduling plans of ammonia synthesis. And then, the pricing model (\ref{eq:pricing_opt}) is solved to obtain the inner electricity and hydrogen price, as well as power distribution between AEHS and AS parts. Finally, the economic assessment is achieved.}

\vspace{-12pt}
\section{Case Studies}
\label{sec:case}
In this section, case studies are performed using the data of a real-life system in Inner Mongolia. First, four benchmarking scenarios are discussed, which reveal the significance of all components in the system. Second, the planning results of the proposed scenarios are displayed, including the analysis of typical weekly and daily operation modes. \textcolor{black}{Then, the robust planning results based on IGDT are discussed.} Finally, a sensitivity analysis of ammonia flexibility is presented.

\vspace{-12pt}
\subsection{Case Description and Setup}
\label{sec:descrip}

To study the proposed method, the planning, operation and trading model proposed in Section \ref{sec:sa_model} and \ref{sec:MIE_model}, and the two-stage decomposed sizing and pricing model proposed in Section \ref{sec:two_stage_model} are established in \emph{Mathematica 12.3} and solved by \emph{Gurobi 9.5.0}.

The real-life system \cite{Batou-Project-2021} located in Baotou City, Inner Mongolia, China,  is used in the case studies, shown in Fig. \ref{fig:OperationTopu}. The wind and solar power curves represent the real historical data from the project. The full load hours (FLH) of wind power are 3500 hours and those of solar power are 1800 hours. According to the actual policies in Inner Mongolia, the maximum net on-grid rate is $r_{\rm{net}} = 20\%$ \cite{MX-Ongrid-rate-2020}.
The nominal annual output of ammonia is given as $\bar{m}_{\mathrm{NH_3}}=10^5$ t. Moreover, the investment, price and operation parameters are listed in Tables \ref{tab:invest_para}, \ref{tab:price_para} and \ref{tab:opera_para}, respectively.

\begin{table}[t]\scriptsize
  \renewcommand{\arraystretch}{2.0}
  \caption{The Investment Parameters \cite{li2021co,gu2022techno}}
  \label{tab:invest_para}
  \centering
  \vspace{-9pt}
  \begin{tabular}{cccccc}
  \hline \hline
  Facility
  &\tabincell{c}{Unit investment cost}
  &\tabincell{c}{O$\&$M cost}
  &\tabincell{c}{Lifetime(years)} \\
  \hline
  WT   &$6000$ RMB/kW    &$2\%$   &$20$ \\
  PV   &$4000$ RMB/kW    &$2\%$    &$20$ \\
  AE   &$3000$ RMB/kW   &$3\%$    &$15$ \\
  HS   &$250$ RMB/$\mathrm{Nm^3}$    &$2\%$   &$15$ \\
  AS   &\tabincell{c}{$0.33$ billion RMB \\($10^5$ t/year)}     &$3\%$    &$15$ \\
  \hline \hline
  \end{tabular}
  \vspace{-9pt}
\end{table}

\begin{table}[t]\scriptsize
  \renewcommand{\arraystretch}{2.0}
  \caption{The Price Parameters}
  \label{tab:price_para}
  \centering
  \vspace{-9pt}
  \begin{tabular}{ccc}
  \hline \hline
  \tabincell{c}{Feed-in-tariff \\$p_{\rm{FIT}}$}
  &\tabincell{c}{Electricity price \\$p_{\rm{purch}}$}
  &\tabincell{c}{Ammonia price \\$p_{\rm{NH_3}}$}\\
  \hline
  $0.2829$ RMB/kWh \cite{MX-Ongrid-rate-2020}     &$0.4572$  RMB/kWh\cite{MX-ElePrice-2020}    &$3200$ RMB/t\cite{CEIC-Economic-Database} \\
  \hline \hline
  \end{tabular}
  \vspace{-12pt}
\end{table}

\begin{table}[t]\scriptsize
  \renewcommand{\arraystretch}{2.0}
  \caption{The Operation Parameters}
  \label{tab:opera_para}
  \centering
  \vspace{-9pt}
  \resizebox{.95\columnwidth}{!}{
  \begin{tabular}{cccccc}
  \hline \hline
  \tabincell{c}{Parameter}
  &\tabincell{c}{Value}
  &\tabincell{c}{Parameter}
  &\tabincell{c}{Value}\\
  \hline
  $N/ \Delta T$ &8760/1h & $\Delta T_{\mathrm{AS}} / T_{\mathrm{trans}}$ &24h/2h \\
  ${\kappa _{{{\rm{H}}_{\rm{2}}}}}$    &$5$ kWh/Nm$^3$\cite{li2021co}     &${\eta _{{\rm{AE,min}}}}/{\eta _{{\rm{AE,max}}}}$   &$5\%/120\%$ load  \\
  ${\kappa _{{\rm{N}}{{\rm{H}}_{\rm{3}}}}}/{\kappa _{{\rm{N}}{{\rm{2}}}}}$   &$0.64/0.24$ MWh/t\cite{klyapovskiy2021optimal}  &${\eta _{{\rm{AS,min}}}}/{\eta _{{\rm{AS,max}}}}$   &$30\%/110\%$ load  \\
  $r_+/r_-$     &$20\%/20\%$ load/h\cite{klyapovskiy2021optimal}    &${\eta _{{\rm{HS,min}}}}/{\eta _{{\rm{HS,max}}}}$   &$10\%/90\%$ load \\
  \hline \hline
  \end{tabular}}
  \vspace{-12pt}
\end{table}

\vspace{-12pt}
\subsection{Benchmarking Scenarios and Planning Results}
\label{sec:benchmarking_results}

Four benchmarking scenarios (denoted as BS) are used for comparison:

{\bf BS1:} Wind-Hydrogen-Ammonia system with a buffer tank for hydrogen storage.

{\bf BS2:} Solar-Hydrogen-Ammonia system with a buffer tank for hydrogen storage.

{\bf BS3:} Wind-Solar-Hydrogen-Ammonia system without a buffer tank for hydrogen storage.

{\bf BS4:} Wind-Solar-Hydrogen-Ammonia system with a buffer tank for hydrogen storage, but the capacities of some components are given: $C_{\rm{W}}=200$ MW, $C_{\rm{S}}=260$ MW and $C_{\rm{AE}}=125$ MW.

The planning results and performance indices are listed in Table \ref{tab:bs_bm_pm}.

Specifically, in BS1, due to the strong volatility of wind power, the revenue is slightly less than that of the proposed scenario, approximately a 2.15\% decrease. However, the difference in the earnings ratio is just 0.46\%.

In BS2, due to the intermittence of solar power, much more off-grid energy is required, i.e., nearly twice the amount of energy compared to BS1, to ensure the stability of the hydrogen supply in ammonia synthesis. At the same time, the capacity of the electrolyzers is 255 MW, leading to the lowest FLH of electrolyzers and earnings ratio, which are only 3875 hours and -21\%, respectively.

In BS3, due to the lack of a buffer tank for hydrogen storage, flexible hydrogen production cannot provide a relatively smooth hydrogen supply for ammonia synthesis. Thus, the volatility of renewable power is almost totally regulated by the external grid, resulting in 17.9\% off-grid energy, while the FLH of electrolyzers exceeds 8000 hours.

In BS4, the complementary characteristics between wind and solar power are not fully considered, and the rate of net on-grid energy is only 9.92\%, which is well below the upper limit (20\%), resulting in a low earnings ratio (0.03\%).

The proposed scenario fully considers the complementary characteristics of wind and solar power, as well as the buffer function of hydrogen storage; therefore, a maximum earnings ratio of 8.15\% is obtained.

\begin{table*}[t]\scriptsize
  \renewcommand{\arraystretch}{2.0}
  \caption{Comparison of Performance Indices of the BSs and the Proposed Scenario}
  \label{tab:bs_bm_pm}
  \centering
  \vspace{-9pt}
  \resizebox{1.9\columnwidth}{!}{
  \begin{threeparttable}[b]
  \begin{tabular}{cccccccc}
  \hline \hline
  Method
  &\tabincell{c}{Optimal size\tnote{a}\\(MW, MW, MW, $\rm{Nm^3}$)}
  &\tabincell{c}{Optimal price\tnote{b}\\(RMB/kWh, RMB/$\rm{Nm^3}$)}
  &\tabincell{c}{Rates of energy \\ exchanging with grid\tnote{c}\\ and curtailment}
  &\tabincell{c}{FLH of electrolyzers\\(hours)}
  &\tabincell{c}{$DTR$\\($10^4$ RMB/year)}
  &\tabincell{c}{Earnings Ratio}
  &\tabincell{c}{CPU times\\(s)} \\
  \hline

  BS1   &$\left\{ 376, \bm{0},160,2.59 \times 10^5 \right\}$       &$\left\{ 0.1898,1.37 \right\}$   &$\left\{ 24.3\%,4.3\%,20.0\%,0\% \right\}$         &$6176$          &$2502.3$     &$7.69\%$   &$ \sim 250$ \\
  BS2   &$\left\{ \bm{0}, 731 ,\bm{255}, 3.77 \times 10^5 \right\}$  &$\left\{ 0.1658,1.37 \right\}$   &$\left\{ 28.7\%,\bm{8.7\%},20.0\%,0\% \right\}$    &$\bm{3875}$          &$\bm{-8419.3}$    &$-21.0\%$  &$ \sim 250$ \\
  BS3   &$\left\{ 356, 38,120,\bm{0} \right\}$                     &$\left\{ 0.1616,1.38 \right\}$   &$\left\{ 37.9\%,\bm{17.9\%},20.0\%,0\% \right\}$        &$\bm{8234}$          &$1147.8$     &$3.83\%$   &$ \sim 200$ \\
  BS4   &$\left\{ 200, 260,125,9.5 \times 10^4 \right\}$           &$\left\{ 0.1840,1.36 \right\}$    &$\left\{ 23.7\%,13.8\%,\bm{9.92\%},0\% \right\}$  &$7905$          &$7.544$      &$0.03\%$   &$ \sim 100$ \\
  Proposed   &$\left\{ 347, 56,155,2.35 \times 10^5 \right\}$      &$\left\{ 0.1966,1.37 \right\}$    &$\left\{ 23.0\%,3.0\%,20.0\%,0\% \right\}$        &$6375$          &$2557.2$     &$\bm{8.15\%}$   &$ \sim 300$ \\
  \hline \hline
  \end{tabular}

  \begin{tablenotes}
  \footnotesize
  \item[a] Optimal size of WTs, PV, AE and HS, respectively.
  \item[b] Optimal price of inner electricity and hydrogen price, respectively.
  \item[c] Rate of on-grid energy, off-grid energy and net on-grid energy, respectively.
  \end{tablenotes}
  \vspace{-12pt}

  \end{threeparttable}
  }

\end{table*}

\vspace{-12pt}
\subsection{The Operation Simulation Analysis in Proposed Scenario}
\label{sec:pm}

\begin{figure}[t]
  \centering
  %\vspace{-12pt}
  \includegraphics[width=3.46in]{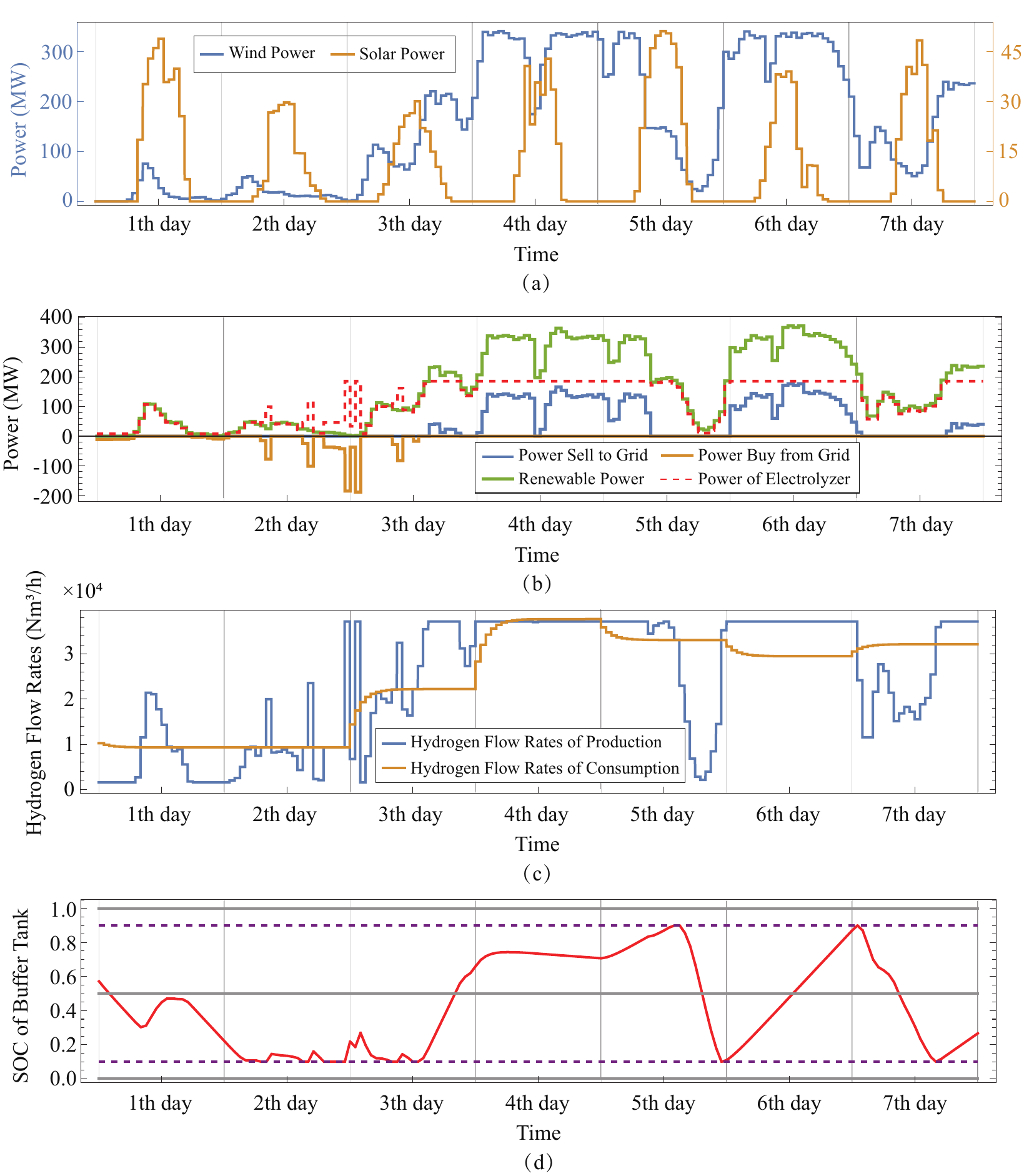}
  \caption{Typical weekly operation in the proposed scenario for optimal sizing and pricing. (a) Renewable power scenarios. (b) Electricity power curves. (c) Hydrogen flow rates. (d) SOC of the buffer tank.}
  \label{fig:TypicalWeek}
  \vspace{-12pt}
\end{figure}

\begin{figure}[t]
  \centering
  %\vspace{-12pt}
  \includegraphics[width=3.46in]{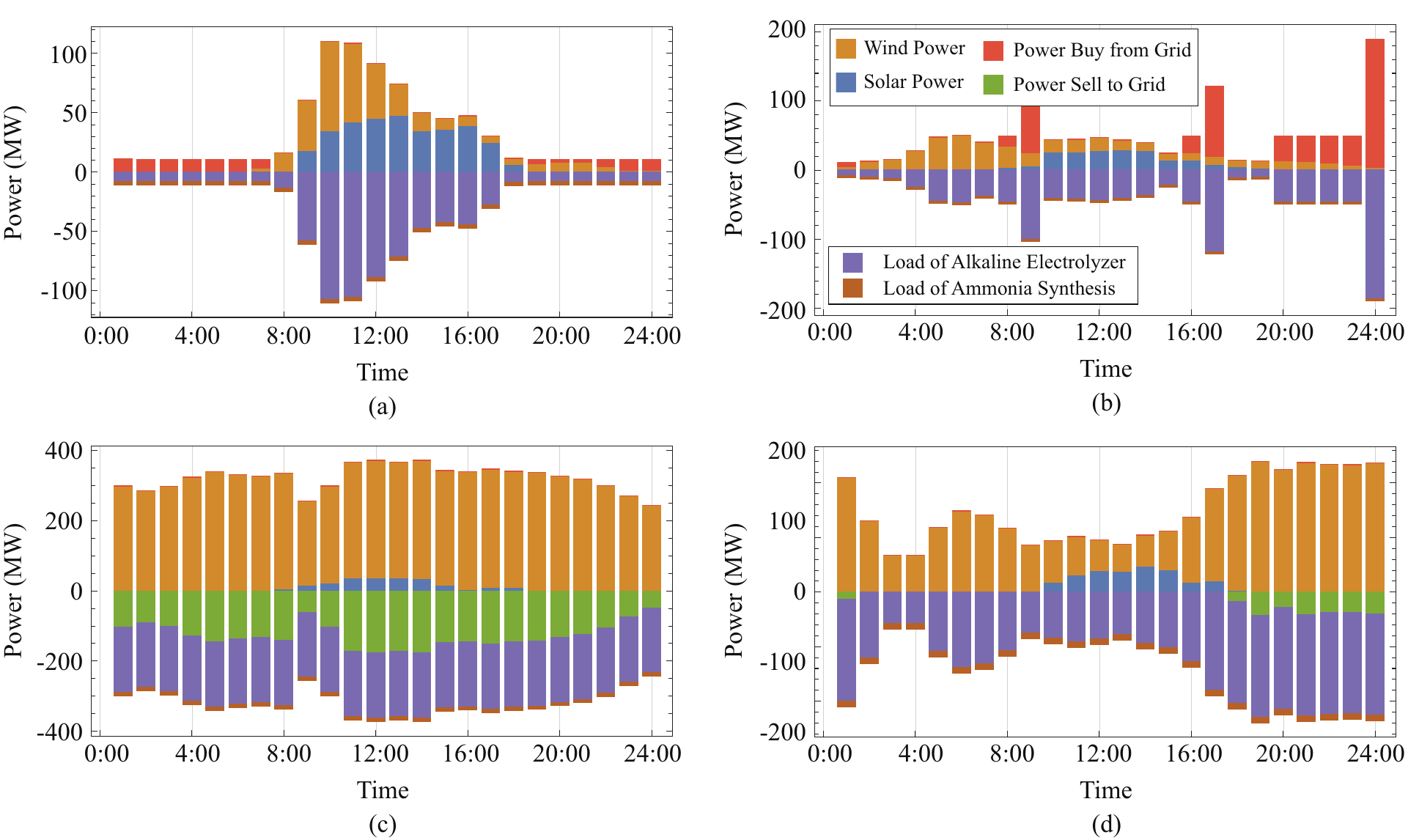}
  \caption{Typical daily power balance in the proposed scenario for optimal sizing and pricing. (a) Small wind, big solar. (b) Small wind, small solar. (c) Big wind, big solar. (d) Wind-solar complementary.}
  \label{fig:TypicalDays}
  \vspace{-12pt}
\end{figure}

By numerical simulation, the optimal operation mode is obtained, and a typical weekly operation mode is plotted in Fig. \ref{fig:TypicalWeek}. Furthermore, four typical daily operation modes in this week are selected to demonstrate the hourly power balance in the daily scenario in detail, as shown in Fig. \ref{fig:TypicalDays}. Under different renewable generation scenarios, there are different operation modes, demonstrated as follows.

\emph{a) Continuous low renewable generation scenario: } From Fig. \ref{fig:TypicalWeek} (a), renewable power is lower than the maximum load of AE and AS almost during the entirety of the 1st-2nd day. In this situation, the load of the electrolyzer almost follows the tracks of renewable power, and the working condition of ammonia is scheduled at a low level, as shown by the orange curves plotted in Fig. \ref{fig:TypicalWeek} (c). Hydrogen stored in the buffer tank is used until it meets the lower limit shown in Fig. \ref{fig:TypicalWeek} (d). However, the power balance of the 1st day and the 2nd day is different. Comparing Fig. \ref{fig:TypicalDays} (a) and Fig. \ref{fig:TypicalDays} (b), due to the initial hydrogen stored in the buffer tank and renewable power generation from 8:00 to 16:00 on the 1st day, there is enough hydrogen that can be supplied to AS. Thus, a slight amount of energy is purchased to guarantee minimum load requirements for hydrogen production, as shown in Fig. \ref{fig:TypicalDays} (a). In contrast, much more energy is purchased on the 2nd day, as shown in Fig. \ref{fig:TypicalDays} (b), because of the lower renewable power generation, and little hydrogen is stored in the buffer tank.

\emph{b) Continuous high renewable generation scenario: } From Fig. \ref{fig:TypicalWeek} (a), renewable power is higher than the maximum load of AE and AS for the entirety of the 4th and 5th days, as well as part time in the 5th day. In this situation, the electrolyzer is at almost full load, even overshooting 120\% of the load. The working condition of ammonia is scheduled at a high level, while the hydrogen storage in the buffer tank continues to increase. Thus, a large amount of energy is sold to the grid, as shown by the green bar plotted in Fig. \ref{fig:TypicalDays} (c).

\emph{c) Normal renewable generation scenario: } From Fig. \ref{fig:TypicalWeek} (a), renewable power is near the maximum load of AE and AS on the 7th day; furthermore, wind and solar power complement each other from 9:00 to 16:00. In this situation, the electrolyzer almost follows the tracks of the renewable power similarly to situation \emph{a)}, sometimes to a full load. Therefore, the system can nearly run in an isolated mode, as shown in Fig. \ref{fig:TypicalDays} (d).

\vspace{-12pt}
\subsection{Robust Planning Result: Optimal Sizing for $\beta>0$}
\label{sec:roubust_planning}
\begin{figure}[t]
  \centering
  %\vspace{-12pt}
  \includegraphics[width=3.46in]{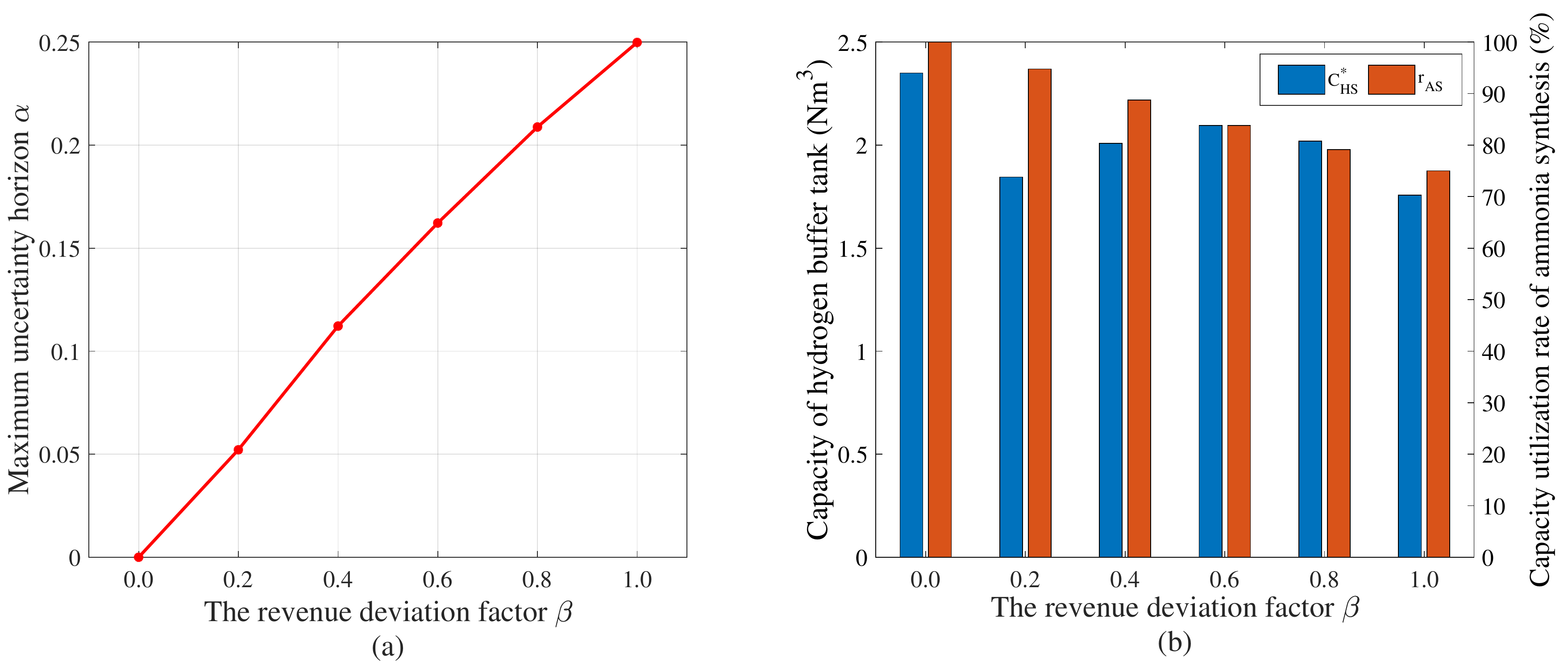}
  \caption{The robust results under different revenue deviation factors. (a) Maximum uncertainty horizon; (b) Robust capacity of HS and related capacity utilization rate of AS.}
  \label{fig:Case_IGDT}
  \vspace{-12pt}
\end{figure}

\textcolor{black}{The numerical simulation results of IGDT-based model (\ref{eq:IGDT_1}) are presented in Fig. \ref{fig:Case_IGDT}, when the revenue deviation factor $\beta$ is set from 0 to 1. When $\beta$ is increasing, the maximum uncertainty horizon $\alpha$ increases, while the capacity utilization rate of AS (defined as $r_{\rm{AS}} = m_{\mathrm{NH_3}}/\bar{m}_{\mathrm{NH_3}}$) decreases, shown in Fig. \ref{fig:Case_IGDT} (a) and (b) respectively. For example, when the uncertainty horizon of renewable power is about 0.05 (deviation within 5\%), the whole system revenue can be guaranteed no less than 80\% of $DTR$ (revenue without uncertainty), while capacity utilization rate of AS decreases from 100\% to 94.78\%.}

\textcolor{black}{Therefore, the proposed method in planning and operation of RePtA system is robust to handle the uncertainty of renewable power, by flexibly adjusting the production of hydrogen and ammonia.}

\vspace{-12pt}
\subsection{Sensitivity Analysis of Ammonia Flexibility}
\label{sec:sensitivity_analysis}

\begin{figure}[t]
  \centering
  %\vspace{-12pt}
  \includegraphics[width=3.4in]{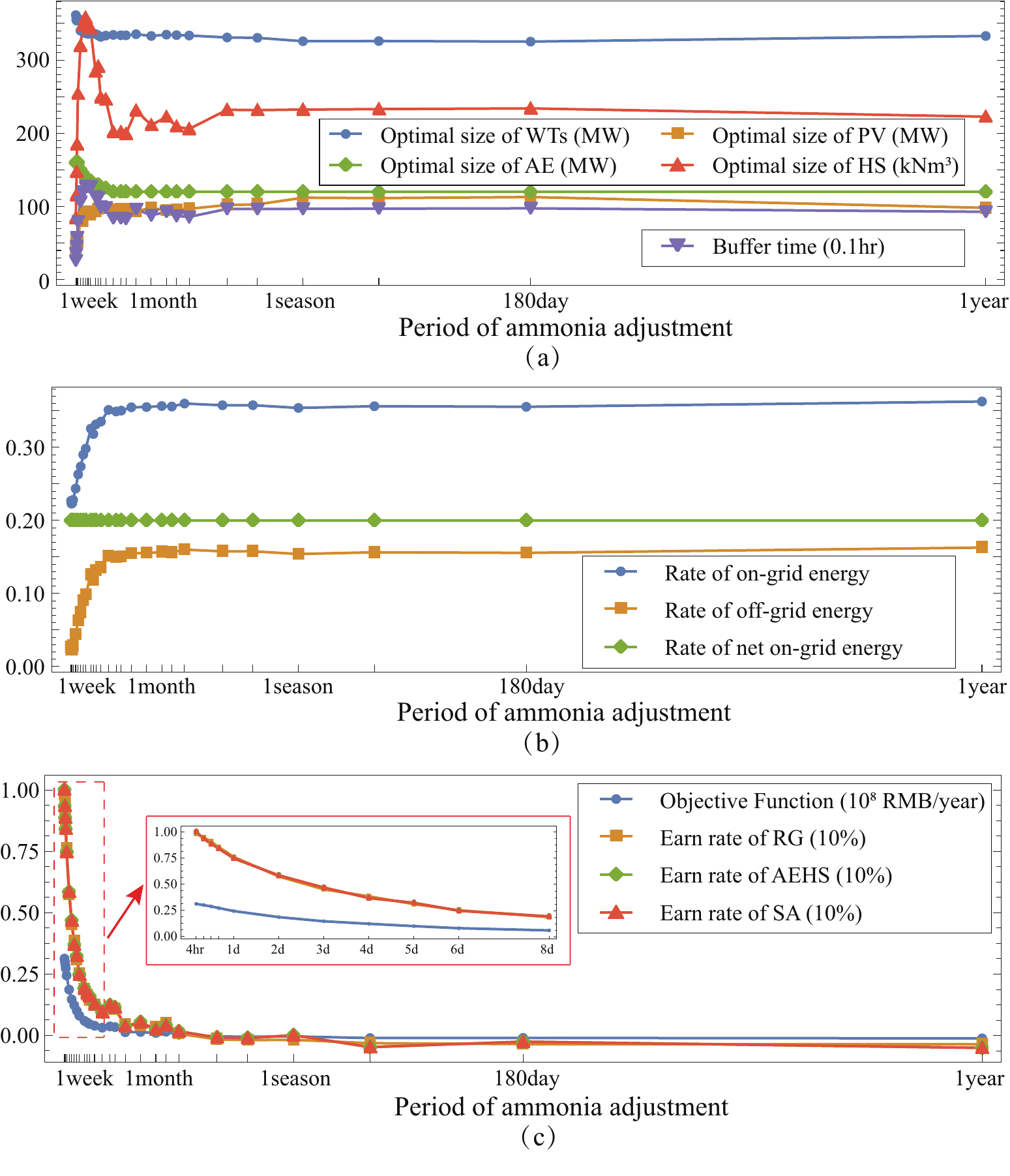}
  \caption{Sensitivity analysis of ammonia flexibility. (a) Result of optimal size. (b) Result of exchange rate with grid. (c) Result of system revenue.}
  \label{fig:periodNH3_Sensitivity_Analysis}
  \vspace{-12pt}
\end{figure}

The flexibility of ammonia is set from no flexibility ($\Delta T_{\mathrm{AS}}=1 $year) to high flexibility ($\Delta T_{\mathrm{AS}}=4 $hours), optimal sizes are shown in Fig. \ref{fig:periodNH3_Sensitivity_Analysis} (a), energy exchange with the grid is shown in \ref{fig:periodNH3_Sensitivity_Analysis} (b), and economic related results are shown in \ref{fig:periodNH3_Sensitivity_Analysis} (c). From the overall trend, with the improvement in the flexibility of ammonia synthesis, the capacity of WTs and electrolyzers increases, while the capacity of PVs decreases. The rate of off-grid energy decreases, which means a decrease in regulation intensity by the power grid. The economics of the system also improves, especially when $\Delta T_{\mathrm{AS}}  \textless 1 $ weeks, as shown by the plots inside the red box in Fig. \ref{fig:periodNH3_Sensitivity_Analysis} (c).

\begin{figure}[t]
  \centering
  %\vspace{-3pt}
  \includegraphics[width=3.5in]{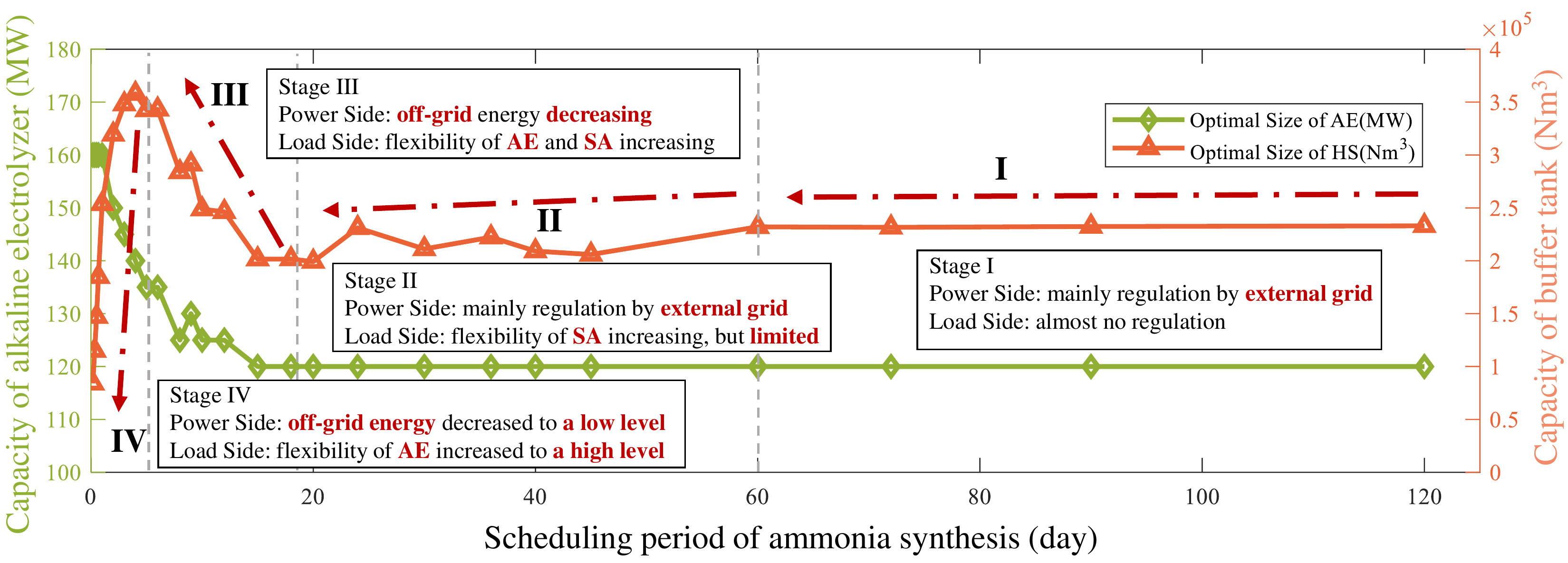}
  \caption{Allocation analysis of buffer tank under different flexibility of ammonia synthesis.}
  \label{fig:Buffer_Tank_Analysis}
  \vspace{-12pt}
\end{figure}

To further discuss the allocation of the capacity of the hydrogen buffer tank, we focus on the capacity of the electrolyzer and buffer tank, as shown in Fig. \ref{fig:Buffer_Tank_Analysis}. It is roughly divided into four stages:

\emph{In stage I}, the capacity of the electrolyzer and buffer tank are fixed, without being affected by the increase in ammonia flexibility. This is because when $\Delta T_{\mathrm{AS}}\textless 2$ months, the flexibility between hydrogen and ammonia cannot be matched. Thus, the fluctuation of renewable power is mainly regulated by the power grid.

\emph{In stage II}, the capacity of the electrolyzer is also fixed, but that of the buffer tank is changing. Although the flexibility of ammonia is increasing, it is still limited, as is the small capacity of the electrolyzer. Thus, the variation range of the buffer tank is also limited.

\emph{In stage III}, when $\Delta T_{\mathrm{AS}}\textless 20$ days, both the capacity of the electrolyzer and buffer tank are increasing. In this situation, the fluctuation of renewable power is transferred into the buffer tank by the load response of AE. The more flexible the regulation of AE is, the larger the capacity of the buffer tank that is needed.

\emph{In stage IV}, when $\Delta T_{\mathrm{AS}} \textless 1$ week, in contrast, the capacity of the buffer tank decreases. This is because the flexibility of ammonia has increased to a relatively high level, which can be well matched to that of AE. Thus, part of the fluctuation in the buffer tank can be regulated by AS.

Although the improvement of the flexibility of ammonia is beneficial to the economics of the system, the cost of transformation flexibility and the security in operation need to be well considered. Therefore, $\Delta T_{\mathrm{AS}}$ between $1\mathrm{day}$ and $1\mathrm{week}$ is a relatively good choice. $\Delta T_{\mathrm{AS}}=1 $day can be used in case studies to show the advanced flexibility of ammonia.

\vspace{-6pt}
\section{Conclusions}
\label{sec:conclusions}

This paper proposes a general scheme of large-scale green hydrogen substitution in the ammonia industry, solved by a two-stage decomposed sizing and pricing optimization method. In the proposed method, first, PtA is modeled as a limited flexible load, and the interest demand of multiple investors is considered. Then, the optimal sizes of all facilities are given by optimization in stage I. Finally, the earnings ratio of different investors is balanced by optimization in stage II, and both the inner electricity and hydrogen price are determined at the same time. Using real data from a real-life system in Inner Mongolia, the planning results are analyzed, and the following conclusions are drawn:

1) Capacity optimization of all facilities is the best choice, with an earnings ratio of 8.15\%, which can fully make use of the complementary characteristics of wind and solar power, as well as make HS play to the greatest extent a regulating role between AE and AS.

2) Sizing before pricing is a reasonable and efficient order in the proposed method; otherwise, the optimal revenue of the system cannot be guaranteed.

3) The flexibility of ammonia plays an important role in affecting the techno-economic effects of RePtA systems, and the scheduling period %in 1 day $\sim$ 1 week
in the range of one day to one week is an appropriate choice to balance the flexibility and safety of ammonia synthesis.

Currently, the electricity price and ammonia price mentioned in this paper are fixed. In future studies, how to participate in electricity, hydrogen and ammonia markets concurrently is a promising work direction.

\bibliographystyle{IEEEtran}
\bibliography{IEEEabrv,WPHSA_OptCapa}

\end{document}